\documentclass[11pt,letterpaper]{article}
\pdfoutput=1 

\usepackage{jheppub} 
\usepackage{tikz}
\usetikzlibrary{shapes,backgrounds}
\usepackage{amscd}

\preprint{DAMTP-2018-04-09}

\title{Homological classification of topological terms in sigma models on homogeneous spaces}

\author[a]{Joe Davighi}
\author[b]{and Ben Gripaios}

\affiliation[a]{Department of Applied Mathematics and Theoretical Physics, University of Cambridge, Wilberforce Road, Cambridge, UK}
\affiliation[b]{Cavendish Laboratory, University of Cambridge, J.~J.~Thomson Ave, Cambridge, UK}
\emailAdd{jed60@cam.ac.uk}
\emailAdd{gripaios@hep.phy.cam.ac.uk}

\abstract{We classify the topological terms (in a sense to be made precise) that may appear in a non-linear sigma model based on maps from an arbitrary worldvolume manifold to a homogeneous space $G/H$ (where $G$ is an arbitrary Lie group and $H \subset G$). We derive a new condition for $G$-invariance of topological terms, which is necessary and sufficient (at least when $G$ is connected), and discuss a variety of examples in quantum mechanics and quantum field theory. In the present work we discuss only terms that may be written in terms of (possibly only locally-defined) differential forms on $G/H$, leading to an action that is manifestly local. Such terms come in one of two types, with prototypical quantum-mechanical examples given by the Aharonov-Bohm effect and the Dirac monopole. The classification is based on the observation that, for topological terms, the maps from the worldvolume to $G/H$ may be replaced by singular homology cycles on $G/H$. In a forthcoming paper we apply the results to phenomenological models in which the Higgs boson is composite.}

\begin{document} 
\maketitle
\flushbottom

\section{Introduction \label{introduction}} 
A sigma model on a homogeneous space is a quantum field theory whose degrees of freedom are maps from some $p$-dimensional worldvolume $\Sigma^p$ to a target space which is a homogeneous space $G/H$ and whose dynamics is described by a $G$-invariant action. Such models are ubiquitous in physics. As examples, in $p=1$ we find many exactly-solvable models in quantum mechanics \cite{Gripaios:2015pfa} (e.g. particles moving in uniform magnetic fields and rigid bodies), while $p \geq 3$ covers all cases in which only some subgroup $H$ of a dynamical symmetry group $G$ is linearly realized {\em in vacuo}, leading to the appearance of Goldstone bosons in the low-energy effective theory. These find applications in particle physics (e.g. the chiral lagrangian \cite{Weinberg:1966fm} and composite Higgs models \cite{Kaplan:1983sm}), condensed matter (e.g. fluids \cite{Gripaios:2014yha} and superfluids \cite{Nicolis:2013lma}) and cosmology (e.g. galileons \cite{Nicolis:2008in}). In between, in $p=2$, we find many interesting examples of conformal field theories and string theories. In short, such sigma models are everywhere. 

Given a system with such degrees of freedom, it is desirable to find all of the possible $G$-invariant terms that may appear in the action. One construction of such action terms was given by Callan, Coleman, Wess, and Zumino \cite{Callan:1969sn}. It requires additional structure in the form of a metric on the worldvolume together with a $G$-invariant metric on $G/H$.\footnote{We note in passing that such a $G$-invariant metric does not necessarily exist.} But it has been known for some time that there exist other $G$-invariant terms, which are topological in
the vague sense that they do not require either metric.

Our goal here is to provide a more-or-less rigorous classification of such topological terms. In order to achieve this goal, we shall need to 
make some assumptions. In the present work, the assumptions, in a nutshell, are that the degrees of freedom of the theory can be replaced by $p$-cycles
on $G/H$ and that the action is obtained by 
integrating (possibly only locally-defined) differential forms on $G/H$ over the cycles. The first assumption allows us to use the power of de Rham's theorem in the classification, whilst the second one guarantees that the action is manifestly local.

These assumptions (which we shall make precise in \S \ref{assumptions}) are strong ones, but they nevertheless lead to a classification which includes many of the known topological terms. The terms that result are topological in the precise sense that the only additional structure required to define them is that of an orientation on the worldvolume.

Ultimately, we find that the possible terms come in one of two types, whose names hark back to prototypical examples.
The first type of terms, which we call Aharonov-Bohm (AB) terms,\footnote{The epithet `theta terms' is common in the literature.} are classified by $H^p(G/H,U(1))$, the $p$th singular cohomology of $G/H$ valued in $U(1)$. Throughout this paper, we shall only consider AB terms corresponding to the free part of $H^p(G/H,U(1))$, which is simply a subquotient of the $p$th de Rham cohomology.
The action for a given term is obtained straightforwardly by integrating any $p$-form in the given de Rham cohomology class on $p$-cycles.
The need to take a quotient accounts for the fact that not all such forms lead to distinct values for the integrand of the path integral, which we write, in unconventional units where $h=1$, as $e^{2\pi i S}$ and which we call the {\em action phase}. Specifically, we take the quotient by the {\em integral} classes (defined as those whose integral over every cycle is an integer). At least when $G$ is connected, $G$-invariance of the action is automatic.

The second type of terms, which we call Wess-Zumino (WZ) terms, are classified by a subspace of the closed, integral $(p+1)$-forms on $G/H$.
The action is not so straightforward in this case, requiring us to integrate locally-defined forms of degree $p,p-1,\dots,0$ (which are constructed from the original $(p+1)$-form, call it $\omega$, via \v Cech cohomology) over $p,p-1,\dots,0$-chains (which are constructed from the original $p$-cycle by subdividing and taking boundaries). The need to take a subspace arises from the requirement that the action be $G$-invariant. Expressed at the level of the globally-defined $(p+1)$-form $\omega$, this requirement turns out to be rather subtle, and is one of the main results of this paper: at least when $G$ is connected, we will show that $G$-invariance requires that the closed $p$-forms $\iota_X \omega$ be exact, for all vector fields $X\in \mathfrak{g}$ that generate the $G$ action on $G/H$.
This requirement, which we call the {\em Manton condition} (for reasons that we will explain in \S \ref{QM examples}), is stronger than that which one might na\"{\i}vely have guessed, namely that the $(p+1)$-form be $G$-invariant.

These WZ terms are so called because they include, as a special case, the term of the same name arising in the chiral lagrangian describing low-energy hadronic physics. Many readers will, no doubt, be familiar with the construction of that term given by Witten~\cite{Witten:1983tw}. Thus, in order to orient the reader, and to motivate the need for the formalism we shall develop, we now briefly review that construction.

Witten's construction is based on homotopy arguments, as follows. If the $p$th homotopy group of $G/H$ vanishes, then any worldvolume homeomorphic to a $p$-sphere is the boundary of a $(p+1)$-ball in $G/H$. Then, one can write a topological action as the integral of the closed, integral, globally-defined $(p+1)$-form $\omega$ over this ball. Requiring the $(p+1)$-form to be closed and furthermore integral guarantees that the resulting action phase is independent of the choice of ball (since any two balls bounding the same worldvolume taken with opposite orientation define a cycle, such that the difference in the action is an integer). Finally, requiring $G$-invariance of the $(p+1)$-form guarantees invariance of the action, without any need to worry about the more subtle Manton condition we described above. 

Given the elegance of this construction, and the simplicity of the resulting condition for $G$-invariance, the reader might wonder why one should bother to instead use our more cumbersome construction of a WZ term (in terms of locally-defined forms).

Witten's approach has, in our opinion, two limitations. The first of these is that, because of the use of homotopy arguments, it is valid only for worldvolumes that are homeomorphic to $S^p$. But, it is clear that we might want to consider worldvolumes of other topology. As Witten himself noted \cite{Witten:1983tx}, the dynamics of the chiral lagrangian in the background of a skyrmion requires us to define the theory on $S^{p-1}\times S^1$. Similarly, in condensed matter, we might wish to employ periodic boundary conditions, giving rise to a toroidal topology; in cosmology, we might wish to consider a Universe of non-trivial topology. In fact, if one believes in quantum gravity, one can make a compelling argument that a physical theory should be defined on worldvolumes of {\em arbitrary} topology (subject to the requirement that they admit the necessary structures, such as spin, that are present in nature). To accommodate this, we switch to homology and provide a construction of topological terms that is valid on worldvolumes of arbitrary topology (subject only to the requirement that they admit an orientation, such that we can integrate differential forms).

The second limitation of Witten's approach is that (in the homotopy language of Witten) it works only if the map from the $p$-dimensional worldvolume to the target is homotopic to the constant map. If not, one cannot define a $(p+1)$-ball on which to integrate the $(p+1)$-form. It does not work, for example, for worldlines homeomorphic to $S^1$ on the torus, as we shall see in \S \ref{QM examples}.
Switching to homology already allows a significant generalization of Witten's approach, in that it allows us to consider $p$-cycles that are the boundary of an arbitrary $(p+1)$-chain. Thus, we are free to consider a worldvolume which is not bounded by a ball, but rather by some more general manifold with boundary. Even then, the homological version of Witten's construction only goes through
when the $p$th singular homology of $G/H$ vanishes, such that every $p$-cycle is a boundary. But switching to homology {\em and} allowing locally-defined forms allows us to construct and classify topological terms that can be defined on {\em all} cycles even when $G/H$ has non-vanishing $p$th singular homology, as we describe in \S \ref{consistency}.
In the general case, we will prove that $G$-invariance of WZ terms requires the full Manton condition.

In the special case of our classification where the homological version of Witten's construction goes through, namely when the $p$th homology of $G/H$ vanishes, there are significant simplications: in this case, our classification shows that not only are there no AB terms, but also the Manton condition 
reduces to requiring $G$-invariance of the form. Moreover, we show in \S \ref{consistency} that the WZ term defined in terms of local forms is, in this special case, indeed equal to that prescribed by Witten's construction. In this case, the space of topological terms is classified straightforwardly by the space of closed, integral, $G$-invariant $(p+1)$-forms on $G/H$.

Despite the generality of our classification, 
the reader might be forgiven for thinking that topological terms are the exception rather than the rule, in the sense that they appear only sporadically in sigma models of physical interest. In fact, they occur in {\em all} of the examples of sigma models that we mentioned in the first paragraph of this Introduction ({\em cf., e.g.,} \cite{Son:2002zn,Dubovsky:2011sk,Delacretaz:2014jka,Goon:2012dy}). They are thus, arguably, almost as ubiquitous as sigma models themselves.

The outline of the paper is as follows. In the next Section, we seek to familiarize the reader with AB and WZ terms through a series of (mostly well-known) examples of both types of term from quantum mechanics. These examples shall, taken together, draw out all the important features of our classification. In \S \ref{assumptions},
we discuss the technical assumptions required, along with their physical justification. In \S\S\ref{ab} and \ref{wz} we derive the classification of AB and WZ terms, and describe a number of quantum field theory examples relevant for physics. In \S\ref{compute}, we discuss how one may compute the space of topological terms for a given $G/H$ and compare with earlier partial classifications \cite{DHoker:1994rdl,dijkgraaf1990}. Section~\ref{conclusion} concludes the discussion.  

\section{An invitation: examples from quantum mechanics \label{QM examples}}

For an example of an AB term, consider quantum mechanics (such that $p=1$), with $G=\mathbb{R}$ and $H=\mathbb{Z}$, such that $G/H \cong S^1$. We can think of this as a particle constrained to move on a circle (in $\mathbb{R}^2$, say), with dynamics that is invariant under translations around the circle.\footnote{This set-up also describes the motion of a rigid body in a plane, in which case the AB term can be thought of as assigning anyonic character to the rigid body, in that the wave function acquires an arbitrary phase under a complete rotation.} Now suppose, as in the AB effect, that a solenoid pierces the centre of the circle. The dynamics remains translationally-invariant, and the solenoid couples to the particle via the magnetic vector potential 1-form $A=bdx$ (where $x \in \mathbb{R}/\mathbb{Z},\ x\sim x+1$, provides us with a coordinate on the circle, and $b \in \mathbb{R}$). Avoiding the homological intricacies for now, we may write the topological action term simply as the integral of the pullback of $A$ over the particle's worldline. For a worldline beginning and ending at the same point, but wrapping the circle $n$ times, the action is given by $bn$ and corresponds (after we have multiplied by the factor $2\pi$ in our normalization of the path integrand) to the AB phase acquired by the wavefunction of the particle during its evolution. Shifting $b\rightarrow b+m$ for any integer $m$ doesn't change this phase, and so the couplings $b$ and $b+m$  describe the same physics.\footnote{As a check, in the presence of a kinetic term $\int \frac{dt}{2}\dot{x}^2$ in the action, the spectrum of the Hamiltonian is given by $E_{k}\left(b\right)=\frac{1}{2}\left(k-b\right)^{2}$, $k \in\mathbb{Z}$, which is indeed invariant under $b \rightarrow b + m$.}
Thus, the topological terms for quantum mechanics on $S^1$ are seen to be in one-to-one correspondence with $ \mathbb{R}/\mathbb{Z} \cong U(1) $. The same result is obtained in our homological classification. Moreover, since all 2-forms vanish on $S^1$, we find that there are no WZ terms in this example.\footnote{A variant of this example, in which we replace $\mathbb{R}/\mathbb{Z} \cong SO(2)\cong S^1$ by $O(2)/O(1) \cong S^1$ illustrates the subtleties that can occur when $G$ is disconnected. Invariance under $O(2)$ restricts $2b \in \mathbb{Z}$, such that the space of AB terms is reduced to $\mathbb{Z}/2\mathbb{Z}$.}

The prototypical (although, as we shall see, not the simplest) example of a WZ term in quantum mechanics arises for a particle moving on $S^2$ in the presence of a magnetic monopole at the centre of the sphere. The physics is rotationally-invariant, so we take $G/H=SO(3)/SO(2)$ $\cong S^2$. The electromagnetic field strength is a closed 2-form proportional to the area form on $S^2$ and may be given, in spherical polar co-ordinates, by $F = \frac{B}{4\pi} \sin \theta d\theta \wedge d\phi$. This is the globally-defined form of degree $p+1=2$ that appears in our classification. Since $F$ is not exact, we cannot write it as the exterior derivative of a globally-defined 1-form $A$. At best, we can write it in terms of 1-forms $A_N=\frac{B}{4\pi}(1-\cos\theta)d\phi$ and $A_S=\frac{B}{4\pi}(-1-\cos\theta)d\phi$, which are singular on $S^2$, but locally well-defined on an open cover consisting of sets $U_N$ and $U_S$, excluding the South and North poles respectively. Dirac obtained his famous quantization condition $B \in \mathbb{Z}$ (which is equivalent to requiring that $F$ be an integral form) by 
requiring that
$A_N - A_S = \frac{B}{2\pi}d\phi $ be a well-defined gauge transformation on $U_N \cap U_S \cong S^1\times \mathbb{R}$. To write the action for a worldline that traverses multiple open sets, requires, as noted by Wu \& Yang \cite{Wu:1976qk}, contributions not just from integrating the different 1-forms on segments of the worldline where they are well-defined, but also requires contributions from evaluating 0-forms (corresponding to Dirac's gauge transformations) at points where we switch between 1-forms.

This prototypical example is indicative of the general story for WZ terms. In generalizing, we adapt ideas of Alvarez \cite{Alvarez:1984es} to a rigorous homological context. Thus, we use a {\em good cover} on $S^2$, namely an open cover (containing at least 4 open sets) in which not only the sets themselves, but also their (finite) intersections, are contractible. Rather than integrate on a worldline, we integrate on a 1-cycle which has been sufficiently subdivided that its constituent chains are contained in individual open sets. As we will see, the twin requirements of adding contributions from $0$-forms and the quantization condition arise explicitly from the desire that the action phase be invariant under diffeomorphisms of the worldvolume that preserve orientation, meaning that the definition of the topological term requires only the structure of an orientation on the worldvolume.

The example of the Dirac monopole has two special features which do not generalize to arbitrary $p$ and arbitrary $G/H$.
The first of these is that the coefficient of a WZ term does not have to take integer values, in general, even though the $(p+1)$-form must be integral. For a counterexample, consider what is arguably the simplest example of a WZ term, which arises for a particle moving in a plane. We thus take $G = \mathbb{R}^2$ and $H$ the trivial subgroup, with dynamics that is invariant under translations. A uniform magnetic field perpendicular to the plane corresponds to a closed, translationally-invariant 2-form $F=Bdx\wedge dy$, with $B \in \mathbb{R}$. This form is exact, since we can write it as  $dA$, with $A=Bxdy$. As a result, its integral over any $(p+1)$-cycle is zero by Stokes' theorem and so it is an integral form for all $B \in \mathbb{R}$.\footnote{For an even more trivial example, note that an AB term, which can be thought of as a WZ term with vanishing $(p+1)$-form, will never have a quantized coefficient.} This topological term, when added to the canonical kinetic term for a particle on the plane, yields the Landau levels in quantum mechanics.\footnote{This example also makes it clear that even exact $(p+1)$-forms can lead to topological terms with physical effects and so the classification of WZ terms for general $G/H$ should involve closed forms rather than cohomology classes, as is oft assumed elsewhere.}

The second feature of the Dirac monopole example which does not generalize is as follows. The action (or rather the action phase, which is the physical object) must be $G$-invariant. For $G/H = SO(3)/SO(2)$ it turns out that $SO(3)$-invariance of the 2-form $F$ is enough to guarantee invariance of the action phase. But in general this is a necessary but not sufficient condition; the loophole arises because the action phase for WZ terms cannot, in the general case where there are non-trivial $p$-cycles, be
expressed directly in terms of the $(p+1)$-form appearing in the classification, but rather is expressed in terms of derived, locally-defined $p,p-1, \dots 0$-forms. The upshot is that we need the stronger condition given earlier.

As ever, a simple example, namely quantum mechanics on the torus, serves to illustrate the point. To this end, let us modify our previous $G=\mathbb{R}^2$ example, now setting $H = \mathbb{Z}^2$, such that $G/H \cong T^2$. Explicitly, at the level of co-ordinates, we identify $x\sim x+1$ and $y\sim y+1$. By analogy with the $\mathbb{R}^2$ example, one might think that there exists an $\mathbb{R}^2$-invariant WZ term corresponding to the closed, translationally invariant 2-form $F=Bdx\wedge dy$, provided we choose $B \in \mathbb{Z}$ so that the integral of $F$ over a fundamental cycle on the torus $T^2$ is an integer.

But, in this example, translation invariance of the 2-form is not enough to guarantee a translationally invariant action. To see the problem, consider a cycle representing a worldline that wraps the $y$-direction once at some constant $x=x_0$, on which we may try to use the locally-well-defined vector potential $A=Bxdy$. But integrating this 1-form over the cycle yields action phase $e^{2\pi i B x_0}$, which is not invariant under (all) translations in the $x$-direction. In fact, there is no choice of local 1-forms that yield a translationally-invariant action for all cycles.

The absence of a WZ term in this example is confirmed by our classification, because the stronger (necessary {\em and} sufficient) condition for $G$-invariance is violated: the interior product of the 2-form with the vector field
$a_x\partial_x+a_y\partial_y$ induced on $T^2$ by the action of the Lie algebra is 
 a closed, but not exact form. Indeed, $\iota_{a_x\partial_x+a_y\partial_y} (Bdx \wedge dy) = a_xBdy -a_y Bdx$ (where $\iota$ denotes the interior product), which is not exact on $T^2$ unless $a_x = a_y =0$.

The curious fact that quantum mechanics on the torus does not admit a translationally invariant magnetic field was noticed long ago by Manton~\cite{Manton:1983mq,Manton:1985jm}; we thus call the generalization of the condition for $G$-invariance of the action phase (which we derive in \S \ref{invariance}) to arbitrary $p$ and $G/H$ the {\em Manton condition}.\footnote{We note that, according to our classification, quantum mechanics on the torus does admit topological terms in the form of AB terms (which vanished on $\mathbb{R}^2$), given by integrating any linear combination of the closed forms $dx$ and $dy$ over cycles.} As we shall see in \S \ref{noether currents}, the Manton condition even has consequences at the level of classical physics.

\section{Formalism \label{assumptions}} 

A classification of topological terms requires a concrete mathematical
starting point, which we now describe, and seek to justify. We assert, on very general physical grounds, that we may 
  equip both the worldvolume and the target space with a smooth structure and insist that the maps between them be smooth. Indeed, our experimental apparatus may only be
  set up, and measurements may only be performed, with finite
  precision; the mathematical description of what happens on scales
  beyond this precision is metaphysics rather than physics, and we are
  free to choose it to be as smooth as we like, without loss of generality.

We may also assume, without loss of generality, that $\Sigma^p$ is connected. Indeed, disconnected components of $\Sigma^p$ may be considered as completely decoupled and so to compare actions on them is to stray once more into the realm of metaphysics.

We also assume, now {\em with} loss of generality, that $\Sigma^p$ is orientable and we choose an orientation on it. Doing so allows us, for example, to integrate differential $p$-forms on $\Sigma^p$, to obtain objects that are invariant under the group, $\mathcal{O}$, of orientation-preserving diffeomorphisms of $\Sigma^p$. Thus, such objects require only the existence of an orientation structure on the worldvolume. We define, correspondingly, a topological term as one that requires only this structure and so is invariant under $\mathcal{O}$.\footnote{This is in accordance with Atiyah's axioms for topological quantum field theory\cite{Atiyah1988}. The functorial axiom, which is needed for locality, is automatically satisfied here because the action is expressed in terms of an integral of a local lagrangian (albeit one supported on chains of various degrees).}

In fact, we will {\em not} define our topological terms by integrating $p$-forms on $\Sigma^p$. The reason is that we wish to bring to bear the power of de Rham's theorem, which requires us to integrate not on manifolds, but on smooth singular chains.\footnote{We recall that a smooth singular $p$-simplex on $G/H$ is a
  map from the standard simplex $\Delta^p \subset \mathbb{R}^p$ to $G/H$ that extends to a smooth map in a neighbourhood of $\Delta^p$. A
  $p$-chain (we drop the qualifier `smooth singular' henceforth) is an element of the free Abelian group on (equivalently a formal finite sum of) such
  simplices and one defines a boundary operator $\partial$ on chains
  that lowers $p$ by one and is such that $\partial^2 = 0$. A
  $p$-cycle is a chain without boundary and a
  $p$-boundary is a cycle that bounds some $(p+1)$-chain.}
To enable us to do so, we make one further assumption on $\Sigma^p$, which is that it is
closed ({\em i.e.} compact without boundary). This assumption requires some discussion. Whilst worldvolumes that are not closed are certainly physically reasonable, one finds in many examples that it suffices to work on closed worldvolumes.
In the
path-integral approach to quantum
mechanics (for which $p=1$), for example, one computes the action phase for all
worldlines beginning at some initial point in the target and ending at
some final point. But what is physical is not the action phase, but
rather the relative difference in the action phase between any two
worldlines. So we can formulate things equivalently by fixing one
worldline and appending it to all other worldlines (with its orientation
reversed and smoothing out the endpoints), making closed worldlines that are all orientation-preserving diffeomorphic to $S^1$.

Similarly, when we move to quantum field theory ($p>1$), we often find that the boundary conditions associated to a given physical situation allow us to assume closure.
Consider, for example, a Euclidean quantum field theory living on
the usual $\mathbb{R}^p$, which is
certainly not compact. Nevertheless, the requirement that the
non-topological part of the 
action be finite typically forces the quantum fields living on it to tend
to a common value
`at infinity', so that we can consider the corresponding worldvolume to be a sphere,
$S^p$, with orientation. Alternatively, we may wish to consider quantum dynamics in the background of some topologically stable object such as a soliton, in which case the Euclidean theory may be considered as a product of spheres.
As another example, in doing condensed matter physics we might wish
(e.g. in studying crystals) to employ periodic boundary conditions in
space, in which case the worldvolume may be taken to be an oriented torus, $T^p$.

The upshot of all these assumptions on the worldvolume is that it defines a fundamental class, $[\Sigma^p]$, as follows.
The (connected) worldvolume $\Sigma^p$ has $p$th homology isomorphic to $\mathbb{Z}$ and $[\Sigma^p]$ is defined to be a generator thereof. Now, the fundamental class is $\mathcal{O}$-invariant and so provides us with a natural object on which to try to define an action (phase) for a topological term.

There exists a natural way to define such an action: take a $p$-form on $\Sigma^p$ and integrate it on any fundamental $p$-cycle (that is, a $p$-cycle in the fundamental class). The $p$-form, being a top-degree form, is necessarily closed, and so, by Stokes' theorem, our definition is independent of the choice of cycle, resulting in an action that is well-defined on $[\Sigma^p]$.

Moreover, there is a natural source of suitable forms: we take any $p$-form on $G/H$ (which need not be closed) and pull it back to $\Sigma^p$ via the map $\phi: \Sigma^p \rightarrow G/H$ that defines the field configuration in the quantum field theory. We can, completely equivalently, define the action by instead integrating the original form on $G/H$ on the cycle on $G/H$ that is obtained by pushing-forward a cycle in $[\Sigma^p]$ to a cycle in $G/H$, where the push-forward is defined by taking the maps $\sigma: \Delta^p \rightarrow \Sigma^p$ defining the constituent simplices of the cycle in $[\Sigma^p]$ and composing with the map $\phi$.

We thus arrive at a formulation of the dynamics in terms of $p$-cycles and $p$-forms on $G/H$. We now wish to modify this in 2 ways. The first way amounts to a restriction on the possible dynamics: we insist that the action be defined on {\em all} cycles in $G/H$, not just on the subset that can be obtained via the push-forward map. We insist on this restriction because it allows us to use de Rham's theorem. In particular, we shall make frequent use of the following results:
\begin{gather}\label{dR}
\text{A differential $p$-form has vanishing integral over every}
\left\{
\begin{aligned}
&\text{$p$-chain} \\
&\text{$p$-cycle}\\
&\text{$p$-boundary}
\end{aligned}
\right\}
\text{iff.\ it}
\left\{
\begin{aligned}
&\text{vanishes.} \\
&\text{is exact.}\\
&\text{is closed.}
\end{aligned}
\right\}
\end{gather}
 
The second modification amounts not to a restriction, but rather to a generalization of the dynamics. To wit, we allow the $p$-forms on $G/H$ to be only locally well-defined. That is, a `form' may consist of distinct pieces, each of which is defined only on a single set, $U_\alpha$ say, of an open cover. In doing so, we
must revise our definition of the action, since the definition we just gave cannot be used for cycles in $G/H$  that intersect multiple open sets. A way forward is found by using the subdivision operator (a standard object in algebraic topology \cite{vick1994homology}) to replace the original cycle by a new cycle whose constituent simplices are contained in single sets in the cover. One may then try to define an action by integrating the locally-defined forms on the simplices where they are well-defined, but this leads to an ambiguity in the following way: suppose that subdivision results in a simplex contained in a double intersection $U_\alpha \cap U_\beta$ of sets. Then there exists a choice of locally-defined $p$-forms which we could integrate on the simplex. 

To remove this ambiguity requires a modification of the action, which we shall discuss in detail in \S \ref{wz}. We end up with an action written in terms of integrals of locally-defined 
$p,p-1,p-2,\dots,0$-forms on chains of corresponding degree, in such a way that the action is well-defined on every cycle in $G/H$. Moreover, as we show in \S\ref{wz}, the definition leads to a well-defined action on $[\Sigma^p]$, {\em ergo} an $\mathcal{O}$-invariant of the worldvolume.

Our further assumptions regarding the target space $G/H$ are few. The group $G$ may be an arbitrary Lie group and $H$ any closed subgroup thereof.\footnote{It is then a theorem that $G/H$ admits the structure of a smooth manifold with a smooth transitive action of $G$.} Neither $G$ nor $H$ need be compact or connected, in general, and we will see that there exist plenty of interesting physical examples where these conditions do not hold.\footnote{There are also interesting physical examples where $G$ is modelled on an infinite-dimensional manifold and so is not, strictly speaking, a Lie group. A prototype is given by a perfect fluid, which may be described, both classically \cite{nla.cat-vn2680980}
and quantum mechanically \cite{Gripaios:2014yha}, as a sigma model in which $G$ contains the group of volume-preserving diffeomorphisms of the manifold on which the fluid flows.} Nevertheless, since we have argued that the worldvolume may be taken to be connected and the map $\Sigma^p \rightarrow G/H$ to be smooth, we may freely take $G/H$ to be connected, if we wish.\footnote{Lest there be any confusion, we remark that neither $G$ nor $H$ need be connected, even when $G/H$ is, {\em cf.} $O(n+1)/O(n) \cong S^n$ or $SO(n+1)/O(n) \cong \mathbb{R}P^n$.}

In what follows, we will see that it is possible to arrive at a straightforward condition for invariance of topological terms under the subgroup of $G$ consisting of elements that are continuously connected to the identity. 
The extra conditions that must be imposed for elements of $G$ that are disconnected from the identity are somewhat subtle for both AB and WZ terms. We will therefore assume throughout the paper that $G$ is connected, and postpone our discussion of the case of disconnected $G$ to Apppendix \ref{disconnected}.

We now discuss the two types of terms arising in our classification, beginning with the rather simpler AB terms.

\section{Aharonov-Bohm terms and their classification \label{ab}} 

Since we are defining our action on $p$-cycles, it obviously makes sense to start by considering integrating $p$-forms, albeit only locally-defined ones. It will be helpful to divide our analysis into two cases, namely in which the local $p$-forms are, or are not, closed. The closed case corresponds to the AB terms, which we discuss in this Section; the other case corresponds to the WZ terms, which we discuss in \S \ref{wz}. 

For the AB terms, we shall take the closed $p$-form to be not just locally, but globally-defined. It turns out that this assumption can be made without loss of generality if one neglects torsion terms in the singular $p$-homology of the target space (or indeed if the torsion vanishes). The proof of this claim is technical and requires the formalism of \S \ref{wz}, so we postpone it to Appendix \ref{local aharonov bohm}. We remark that one may incorporate torsion terms into a homological classification of topological action terms through {\em locally}-defined AB terms. If one includes this torsion piece, the full space of AB terms is the group $H^p(G/H,U(1))$, the $p$th singular cohomology of $G/H$ valued in $U(1)$.


Let $A$ be a closed, globally-defined $p$-form on $G/H$. Define a topological action, evaluated on a generic worldvolume $\Sigma^p$, by its integral over 
a $p$-cycle $z$ in $G/H$ which is the push-forward of a cycle in $[\Sigma^p]$:
\begin{equation}
S[z]=\int_{z} A. \label{ab action}
\end{equation}
This integral vanishes for any exact form by (\ref{dR}), and so only depends on the de Rham cohomology class of $A$.
Since any two fundamental cycles differ by a boundary, and any $p$-boundary in the source pushes forward to a $p$-boundary in the target, then by (\ref{dR}) every fundamental cycle yields the same action (\ref{ab action}), because $A$ is closed. Hence, (\ref{ab action}) is well-defined on the fundamental class $[\Sigma^p]$, and is therefore $\mathcal{O}$-invariant.

The action for AB terms has 3 other special properties, none of which will hold for WZ terms. 
The first is that, by the Poincar\'e lemma, an AB term is locally exact; like a total derivative in the lagrangian, it therefore gives no contribution to the classical equations of motion, such that its effects are purely quantum-mechanical. The second is that it gives no contribution to perturbative Feynman diagrams (for $p>2$). The third property is that the AB terms only yield non-trivial action phase when $G/H$ admits $p$-cycles that are not $p$-boundaries, {\em i.e.} when the $p$th homology is non-vanishing.

Having identified the source of AB terms, namely closed, global $p$-forms, we now consider their classification.

\subsection{Classification}

Generally, we shall need to check three things when we classify the possible topological terms, which we refer to as {\em consistency}, {\em invariance}, and {\em injectivity}. In more detail, the notions are as follows:
\begin{itemize}
\item consistency:
we have prescribed that the action must be defined on every $p$-cycle in $G/H$. If we are constructing an action from differential forms that are only locally-defined on open sets, we must ensure there are no ambiguities where sets overlap. Moreover we must check that the action is
well-defined on the fundamental class after we pull back to the source;
\item invariance: the action must be $G$-invariant;
\item injectivity: na\"{\i}vely, every coupling $g$ that appears in an action is just a real number (though consistency may force it to be an integer). But if two different numbers lead to the same value of $e^{2\pi iS}$ on all possible cycles, then the physics will be the same. So we need to check that the space of couplings injects to the space of action phases.
\end{itemize}

We have defined an AB term as the integral of a globally-defined $p$-form over any cycle in $G/H$, and so
there are no ambiguities pertaining to the integration of local forms. We have shown above that such an integral is well-defined on $[\Sigma^p]$, and therefore $\mathcal{O}$-invariant. The integral (\ref{ab action}) thus defines a consistent topological action.
Moreover, an AB term is invariant under $G$, if (as we are assuming for the present purposes) $G$ is connected. To see this, consider an infinitesimal $G$ transformation, generated by vector field $X$ on $G/H$. The action (\ref{ab action}) varies by 
\begin{equation}
\delta_X S[z] = \int_z L_X A = \int_z d\iota_X A = \int_{\partial z} \iota_X A = \int_0  \iota_X A = 0, \label{ab invariance}
\end{equation}
where $L_X$ is the Lie derivative. In the second equality, we applied Cartan's homotopy formula, $L_X=d\iota_X +\iota_X d$, together with $dA=0$, and in the third equality we applied Stokes' theorem. Finally, $\partial z=0$ because $z$ is a cycle.

The vector fields $X$ define, via their integral curves, an action of the subgroup of $G$ given by the image of the exponential map, $\exp : \mathfrak{g} \rightarrow G$. So (\ref{ab invariance}) implies invariance under the action of $\exp(\mathfrak{g})\subset G$. Unfortunately, the exponential map is not surjective in general, even when $G$ is connected. It is, however, a theorem that any element $g$ of a connected group $G$ can be written as a product of a finite number of elements in $\exp(\mathfrak{g})$. Hence, an AB term is invariant under the action of the connected group $G$, for any closed $p$-form $A$.\footnote{For a slicker proof, one may simply note that the action of the connected component of $G$ takes cycles into homologous cycles; it then follows immediately that the AB term is invariant, because the integral of a closed form over a cycle depends only on the homology class of the cycle.}

It remains to check injectivity. To do so, note that
if (and only if) two $p$-forms $A$ and $B$  differ by a form that is integral, {\em i.e.} such that $\int_{z}(A-B)\in\mathbb{Z}$ for any $p$-cycle $z$, then the corresponding action phases $\exp(2\pi i \int_{z} A)$ and $\exp(2\pi i \int_{z} B)$ will agree on all $p$-cycles. We saw this explicitly in the  example of quantum mechanics on $S^1$ in \S \ref{QM examples}. Thus, we must take the quotient 
\begin{equation}
H^{p}_{dR}(M,\mathbb{R})/H^{p}_{dR}(M,\mathbb{Z}),
\end{equation}
of the real de Rham cohomology with respect to its subgroup of integral classes to define the space of physically inequivalent AB terms. 

We remark that the set of inequivalent AB terms thus obtained has the structure of an Abelian Lie group. This is no accident, in that it accords with one's physical expectation that two topological actions can be added (in either order) to make a third action which is also topological, \&{\em c.}, and also that small enough changes in the values of the couplings should be physically indetectable. The same structure will be present on the set of WZ terms, and we will have occasion to exploit it in what follows.

 We now give 3 more examples of AB terms in field theory, namely the $\mathbb{C}P^N$ model and a model exhibiting $T$-duality in $p=2$, and the minimal composite Higgs model in $p=4$, where we point out the existence of a new topological term.

\subsection{Examples}

\subsubsection{The 2-d $\mathbb{C}P^N$ model}

Consider a $p$-dimensional sigma model on complex projective space, $\mathbb{C}P^N$, which may be realised as a homogeneous space with  $G/H = U(N+1)/(U(N)\times U(1))$. Its $p$th homology (with integer coefficients) is given by $\mathbb{Z}$ for $p$ even between $0$ and $2N$, and vanishes otherwise. The corresponding real cohomology groups are equal to $\mathbb{R}$ or $0$.

The model with $p=2$ is well studied in physics, particularly at large $N$, where various simplifications occur ~\cite{DAdda:1978vbw,Coleman:1976uz}. Recall that $\mathbb{C}P^N$ may be parametrized by $N+1$ projective co-ordinates, that is, a set of complex numbers $z_i\in\mathbb{C}$, $i=1,\dots,N+1$, together with the constraint $\sum z_i^* z_i =1$ and the $U(1)$ equivalence $z_i \sim e^{i\alpha} z_i$.
The second de Rham cohomology $H_{dR}^2(\mathbb{C}P^N,\mathbb{R})=\mathbb{R}$ has a single generator, which we can take to be the K\"ahler form, $\frac{i}{2}dz^i \wedge d\bar{z}^i$ in our co-ordinates. Hence, there is an AB term for any choice of $N$, obtained by integrating a 2-form proportional to the K\"ahler form over 2-cycles in $\mathbb{C}P^N$. Taking the quotient by the subgroup of forms that are integral, the space of topological terms is given by $\mathbb{R}/\mathbb{Z} \cong U(1)$.

This term is often called a `theta term' in the literature, because it is a close analogue of the theta term in QCD. Indeed, the constraint $\sum z_i^* z_i =1$ and the equivalence relation $z_i \rightarrow e^{i\alpha} z_i$ can be enforced in  field theory via a lagrange multiplier $\lambda(x)$ and a $U(1)$ gauge field $A(x)$, respectively. If one just has the quadratic kinetic term plus the theta term in the lagrangian, one can integrate out the $z_i$ in the (Gaussian) path integral to obtain a theory involving only the fields $A(x)$ and $\lambda(x)$. If one then takes the large $N$ limit\footnote{$\mathbb{C}P^{\infty}$ plays a special role in mathematics too: it is the Eilenberg-Maclane space $K(\mathbb{Z},2)$.} of the resulting effective lagrangian the theory reduces to that of a dynamical gauge field with the usual theta term of electromagnetism in $p=2$, studied by Schwinger and others as a two-dimensional model for real-world QCD~\cite{1962PhRv..128.2425S}.

\subsubsection{$T$-duality on the torus}

Suppose that $p=2$ and that $G/H = (\mathbb{R}/\mathbb{Z})^2 \cong T^2$. 
Since $H^2_{dR} (T^2, \mathbb{R}) = \mathbb{R}$, our classification indicates that there is an AB term given by the integral of a form proportional to the translationally-invariant volume form on $G/H$. This will result in a non-trivial action phase only when the worldvolume has itself the topology of a 2-torus, so let us suppose that this is the case. We thus have a model with maps from a worldsheet $T^2$ to a target $T^2$ with a topological AB term with values in $\mathbb{R}/\mathbb{Z}$. Adding the usual two-derivative kinetic term results in a model exhibiting $T$-duality, in which the topological term plays a key r\^{o}le, pairing up with the geometric area of the torus to make a complex parameter which gets interchanged under $T$-duality with the complex structure parameter of the torus.

\subsubsection{The 4-d minimal composite Higgs model \label{minimal composite higgs}} 

For a final example, consider the minimal composite Higgs model (MCHM) \cite{Agashe:2004rs} in $p=4$, in which the electroweak hierarchy problem is `solved' by postulating that the Standard Model Higgs field emerges (as a composite pseudo Nambu Goldstone boson) from some strongly-coupled dynamics whose form at high energies is unspecified, but which can be described at low energies by a sigma model with target space $G/H=SO(5)/SO(4) \cong S^4$. Since $H_{dR}^4(S^4,\mathbb{R})=\mathbb{R}$, and $H_{dR}^4(S^4,\mathbb{Z})=\mathbb{Z}$, there is an AB term given by the integral of a 4-form proportional to the volume form on $S^4$. The space of inequivalent topological action phases is thus, yet again, 
$\mathbb{R}/\mathbb{Z}=U(1)$. While a topological term (of WZ type) is known to exist in a non-minimal composite Higgs model with $G/H=SO(6)/SO(5)$ ~\cite{Gripaios:2009pe} (see \S~\ref{beyond minimal composite higgs}), this topological term in the minimal model has not been noticed so far in the literature, to our knowledge.

We postpone a fuller discussion of the resulting phenomenology to~\cite{Davighi:2018xwn}. For now, we remark that, while the low-energy effects are expected to be very small, the fact that the term violates both $P$ and $CP$ may nevertheless lead to interesting effects. We will see more examples of topological terms in composite Higgs models in \S\ref{beyond minimal composite higgs}, when we consider WZ terms.

\section{Wess-Zumino terms and their classification \label{wz}
}

Now we turn to topological terms corresponding to $p$-forms on $G/H$ that are not closed, which we call WZ terms. We begin by remarking that one cannot capture all such terms by requiring the $p$-form $A$ to be globally-defined. Nevertheless, even if $A$ is only locally-defined, consistency demands that $dA$ (which is now non-zero) \textit{is} globally-defined. Perhaps the easiest way to see this is to take the classical limit. One finds that $dA$ appears directly in the classical equations of motion, and so should be well defined everywhere on $G/H$ for the classical limit to exist. Thus, a useful starting point for constructing a WZ term is a globally-defined $(p+1)$-form $\omega$ on $G/H$. Such a form is necessarily closed since, at least locally, $\omega=dA$.

To see the kind of restrictions we will have to place on $\omega$, it is helpful to first consider the special case when $A$ is itself globally-defined, and then return later to the general case. If $A$ is globally-defined, then $\omega$ is exact, and we can define an $\mathcal{O}$-invariant, and thus topological, action simply by integrating $A$ over $p$-cycles.\footnote{Note that, unlike for the Dirac monopole example, there is no quantization condition on the coefficient of $\omega$ in this case, because exactness implies that its integral is zero over any $(p+1)$-cycle (so $\omega$ is automatically an integral form).} In this case, the $p$-form $A$ can be regarded as a lagrangian for the theory, but we shall see that when $A$ is only locally-defined, there is no well-defined notion of the lagrangian. 

To be $G$-invariant, we must require $\int_{z} (L_g^*-1) A=0$ for all $p$-cycles $z$.  By (\ref{dR}), this is true iff.\ $(L_g^*-1) A$ is exact, for all $g$.
In other words, the `lagrangian' $A$ may be `quasi-invariant', in the sense that it shifts by a total derivative under the symmetry.\footnote{We note that, in general, it is meaningless in general to try to define a WZ term, as others have done, as a term in the lagrangian that shifts by a total derivative under the action of $G$; in general, as we have just remarked, there is no lagrangian!} It follows that the $(p+1)$-form $\omega$ is strictly invariant, $(L_g^*-1)\omega = 0$, because the exterior derivative commutes with pullback. As we mentioned in \S \ref{QM examples}, the Landau problem on $\mathbb{R}^2$ is an example of this special case.

Now let's go back to the general case, where $A$ need only be locally-defined. In other words, we
suppose that $\omega$ is not exact. We choose an open cover $\{U_\alpha\}$
of our target space, such that $A$ is well defined on each open set, taking value $A_\alpha$ on $U_\alpha$. Given such a collection $\{A_\alpha\}$ of local $p$-forms, it is no longer obvious, {\em a priori}, how to write down an action phase $e^{2\pi i S[z]}$ for each $p$-cycle $z$ in $G/H$, which is consistent, let alone $G$-invariant. If it is the case that the worlvolume cycle $z$ is in fact a boundary, $z=\partial b$, then one can follow Witten's construction and integrate a $G$-invariant $(p+1)$-form $\omega$ directly over the $(p+1)$-chain $b$ to obtain a manifestly $G$-invariant action \cite{Witten:1983tw}. 
If not, we must deal with local forms directly (and there is certainly no well-defined notion of a lagrangian). We do so, following Wu \& Yang ~\cite{Wu:1976qk} and Alvarez ~\cite{Alvarez:1984es}, by writing a topological action phase in terms of contributions on the open sets in our cover, and on finite intersections thereof. 

We shall again need to make sure that the action phase so defined satisfies our triumvirate of criteria, namely consistency, invariance, and injectivity. For WZ terms, consistency leads to the quantization condition (specifically, the requirement that $\omega$ be an integral $(p+1)$-form), as we explain using tools borrowed from \v Cech cohomology and sheaf theory in \S \ref{consistency}; invariance leads to the Manton condition, which we derive in \S \ref{invariance}; injectivity follows straightforwardly, as we show in \S \ref{injectivity}.
Together, these restrictions define the appropriate subspace of closed $(p+1)$-forms that are, as claimed in the Introduction, in one-to-one correspondence with $G$-invariant WZ terms on $G/H$.

\subsection{Consistency and the quantization condition \label{consistency}} 

We now describe (in a very simplistic way; for more details, see ~\cite{bott1995differential}) the elements of \v Cech cohomology and sheaf theory that we need.

We assign, to each open set $U \subset G/H$, an abelian group $\mathcal{F}(U)$; we will, according to our needs, variously take $\mathcal{F}(U)$ to be the smooth $q$-forms, $\Lambda^q(U)$, on $U$, or constant maps $U \rightarrow \mathbb{R}$, or constant maps $U \rightarrow \mathbb{Z}$.\footnote{The objects $\mathcal{F}$ thus defined are, in fact, examples of {\em presheaves}, but we will sidestep the technicalities here.} Every smooth manifold (and thus every $G/H$) admits a {\em good cover}, $\mathcal{U} = \{U_\alpha \}$, namely an open cover satisfying the additional property that the open sets $U_\alpha$, and all finite intersections (where we define $ U_{\alpha_0} \cap U_{\alpha_1} \cap \dots \cap U_{\alpha_p} := U_{\alpha_0 \alpha_1  \dots  \alpha_p}$)  thereof, are contractible.  For example, $\mathbb{R}^n$ has a good cover with 1 open set, $S^1$ has a good cover with 3 open sets, and $S^2$ has a good cover with 4 open sets.\footnote{Note that the open cover of $S^2$ considered in \S \ref{QM examples} with just two open sets, $U_N=S^2\backslash\{S\}$ and $U_S=S^2\backslash\{N\}$, is not a good cover because the intersection of these two sets is clearly not contractible; rather, a good cover can be formed by projecting the four faces of a tetrahedron onto the sphere, and enlarging them slightly such that they intersect.}  The utility of a good cover is that we can use the Poincar\'e lemma on the open sets and their finite intersections. Given a good cover $\mathcal{U}$ we define a \v Cech $p$-cochain on $\mathcal{U}$ with values in  $\mathcal{F}$ to be an element of the group
\begin{gather}
\check{C}^p (\mathcal{U}, \mathcal{F}) = \sum_{\alpha_0 < \alpha_1 < \dots < \alpha_p} \mathcal{F} (U_{\alpha_0 \alpha_1  \dots  \alpha_p}).
\end{gather}
Thus $\omega \in \check{C}^p (\mathcal{U}, \mathcal{F})$ may be characterized by the set of values $\{\omega_{\alpha_0 \alpha_1  \dots  \alpha_p} \in \mathcal{F} (U_{\alpha_0 \alpha_1  \dots  \alpha_p})\}$ that it takes on the $(p+1)$-fold intersections.\footnote{Hereafter, we will allow the indices specifying the components of a \v Cech cochain to be in any order (not just ascending), subject to the condition that $\omega_{\alpha_0 \alpha_1  \dots  \alpha_p}$ is antisymmetric on all pairs of indices.}
The \v Cech coboundary operator $\delta_p: \check{C}^p (\mathcal{U}, \mathcal{F}) \rightarrow \check{C}^{p+1} (\mathcal{U}, \mathcal{F})$ is defined via its action on $\omega_{\alpha_0 \alpha_1  \dots  \alpha_p}$ by 
\begin{equation}
(\delta\omega)_{\alpha_0 \alpha_1  \dots  \alpha_{p+1}} = \sum_{i=0}^{p+1} (-1)^i \omega_{\alpha_0  \dots \hat{\alpha_i} \dots \alpha_{p+1}},
\end{equation}
where a $\hat{}$ denotes omission of the index, whence one may check that $\delta_p \circ \delta_{p-1} = 0$. We define the $p$th \v Cech cohomology of $G/H$ with values in $\mathcal{F}$, $\check{H} (G/H, \mathcal{F})$ to be the usual cohomology of the complex $\check{C} (\mathcal{U}, \mathcal{F})$, {\em viz.} $\mathrm{ker} \; \delta_p / \mathrm{im} \; \delta_{p-1}$. 

As our notation suggests, the cohomology $\check{H} (G/H, \mathcal{F})$ turns out to be independent of the choice of good cover $\mathcal{U}$. In fact, when we choose $\mathcal{F} (U)$ to be the constant real-valued functions on $U$, we find that the cohomology that results is isomorphic to the usual de Rham cohomology. 

To see the relevance of this mathematical formalism to our physical problem, let us return again to our starting point: we consider a globally-defined, closed (but not necessarily exact), $(p+1)$-form on $G/H$, which we denote by $\omega$. The idea is that $\omega$, provided that it satisfies some additional criteria, can be used to define a topological term.
To see how the term comes about, we first note that $\omega$ defines an element of $\check{C}^{0} (\mathcal{U}, \Lambda^{p+1})$ by restricting $\omega$ to each of the $U_{\alpha}$: $\omega_{\alpha}:=\omega|_{\alpha}$. Using the Poincar\'e lemma, we may then construct an element $\{A^p_{\alpha}\} \in \check{C}^{0} (\mathcal{U}, \Lambda^{p})$ via
\begin{equation}
dA^p_{\alpha}=\omega_{\alpha}, \ \mathrm{on}\ U_{\alpha}. \label{omega}
\end{equation}
Since $\omega$ is globally-defined, we must have that $\omega_{\alpha}=\omega_{\beta}$ on $U_{\alpha\beta}$. Hence $d(A^p_{\alpha}-A^p_{\beta})=0$ and, again by the Poincar\'e lemma, we may construct an element $\{A^{p-1}_{\alpha\beta}\} \in \check{C}^{1} (\mathcal{U}, \Lambda^{p-1})$ via
\begin{equation}
A^p_{\alpha}-A^p_{\beta}=dA^{p-1}_{\alpha\beta}, \quad \mathrm{on}\ U_{\alpha\beta}. \label{transition}
\end{equation}
This set of conditions on double intersections can be expressed concisely using the \v Cech coboundary operator, as 
\begin{equation}
\delta\left\{ A^p_{\alpha}\right\} =\left\{ dA^{p-1}_{\alpha\beta}\right\}.
\end{equation}
We note, moreover, that the \v Cech 0-cochain $\{\omega_{\alpha}\}=\{dA^p_{\alpha}\}$ is in fact a \v Cech co\textit{cycle}: $\delta\{\omega_{\alpha}\}=\{\omega_{\alpha}-\omega_{\beta}\}=0$, because $\omega$ is globally-defined.

Now let us use this formalism to construct a consistent topological action phase for any $p$-cycle $z$ in $G/H$. In order to integrate the $p$-forms $\{A^p_{\alpha}\}$ which are locally-defined on the open sets in $\mathcal{U}=\{U_{\alpha}\}$, the chains on which we are to integrate must be contained within these open sets; such chains are referred to as $\mathcal{U}$-small. Thus, we first apply the {\em subdivision operator}, $\mathrm{Sd}$, as many times, $n$ say, as is necessary (we refer the reader to, {\em e.g.}, \cite{vick1994homology} for details of the construction).
The original cycle $z$ we started with is mapped to a homologous cycle $\mathrm{Sd}^n z$, which is a formal sum of a set of $\mathcal{U}$-small $p$-chains, which we denote $\{c_{p,\alpha}\}$, where $\mathrm{Im}\ c_{p,\alpha}\subset U_{\alpha}$ and such that $\mathrm{Sd}^n z=\sum_{\alpha}c_{p,\alpha}$, on which we can now integrate the local $p$-forms $\{A^p_{\alpha}\}$. 

Having done so, one might na\"\i vely try to define the action to be $S=\sum_{\alpha}\int_{c_{p,\alpha}}A^p_{\alpha}$.
This is not a good definition, however, because there is an ambiguity whenever a particular $p$-simplex is contained not just in an open set $U_{\alpha}$, but rather in the intersection of two open sets, say $U_{\alpha\beta}$. The na\"\i ve action is ambiguous because we could choose to integrate $A^p_{\alpha}$ or $A^p_{\beta}$ on this simplex. To fix this problem, we shall need to add pieces to the action corresponding to integrals over $(p-1)$-chains of the $(p-1)$-forms $A^{p-1}_{\alpha\beta}$ defined in (\ref{transition}) to compensate for the ambiguity. However, such a fix introduces further ambiguities to fix up. 
Rather than fixing up the ambiguities one by one, we shall now cut to the chase and explain from the top down how to construct an action phase from local forms which is ambiguity-free.

It turns out that to construct such an action, one needs not just the local forms $\{A^p_{\alpha}\}$ and $\{A^{p-1}_{\alpha\beta}\}$ that we have so far constructed, but rather a whole tower of locally-defined forms of degree $p,p-1,\dots,0$.
We have already constructed, using the Poincar\'e lemma, an element $\{A^p_{\alpha}\} \in \check{C}^{0} (\mathcal{U}, \Lambda^{p})$ and an element $\{A^{p-1}_{\alpha\beta}\} \in \check{C}^{1} (\mathcal{U}, \Lambda^{p-1})$, which satisfy $\{dA^p_{\alpha}\}=\{\omega_{\alpha}\}$ and $\{dA^{p-1}_{\alpha\beta}\}=\delta\{A^p_{\alpha}\}$. We proceed in a similar way to construct elements $\{A^{p-q}_{\alpha_0 \alpha_1 \dots \alpha_q}\}\in \check{C}^{q} (\mathcal{U}, \Lambda^{p-q})$ (that is, in words, a set of $(p-q)$-forms defined locally on $(q+1)$-fold intersections of the open sets in our good cover) for each $0\leq q\leq p$, which satisfies
\begin{equation}
\{dA^{p-q}_{\alpha_0\dots\alpha_{q-1}\alpha_q}\}=\delta\{A^{p-q+1}_{\alpha_0\dots\alpha_{q-1}}\}. \label{cech relations}
\end{equation}
Using this equation, we can construct each $\{A^{p-q}\}$ from $\{A^{p-q+1}\}$ (where we shall sometimes suppress the indices for clarity) by first applying the \v Cech coboundary operator, and then using the Poincar\'e lemma to ``undo" the exterior derivative $d$. Thus, starting from the local $p$-forms $\{A^p_{\alpha}\}$, we construct a whole tower of locally-defined forms of degree $p,p-1,\dots,0$. 

The \v Cech cochains thus defined are also cochains in the de Rham complex (restricted to open sets and appropriate intersections thereof). In this sense, they sit inside a double cochain complex acted upon by both the exterior derivative $d$ and the \v Cech coboundary operator $\delta$. We can illustrate the consistency relations (\ref{cech relations}) conveniently by gathering the double cochains we have constructed into a \textit{tic-tac-toe} table (see~\cite{bott1995differential} for details), whose $(q,r)$th entry is an element in $\check{C}^r(\mathcal{U},\Lambda^q)$: 
\begin{equation}
\begin{array}{cc|cccccccc}
\Lambda^{p+2} &  & 0\\
\Lambda^{p+1} & \omega & \{ \omega_{\alpha}\}  & 0\\
\Lambda^{p} &  & \{ A^p_{\alpha}\}  & \delta\{ A^p_{\alpha}\}=\{dA^{p-1}_{\alpha\beta}\} \\
\Lambda^{p-1} & \vdots & \vdots & \{A^{p-1}_{\alpha\beta}\} \\
\vdots & \vdots & \vdots & \vdots \\
\Lambda^{1} &  &  &  & \{A^1_{\alpha_{0}...\alpha_{p-1}}\} & \delta\{ A^1_{\alpha_{0}...\alpha_{p-1}}\} & 0 \\
\Lambda^{0} &  &  &  & \ldots & \{ A^0_{\alpha_{0}...\alpha_{p}}\}  & \delta\{ A^0_{\alpha_{0}...\alpha_{p}}\}  & 0\\
\hline
d & \uparrow &  &  & \ldots &  & \{ K_{\alpha_{0}...\alpha_{p+1}}\} \\
\delta & \rightarrow & \check{C}^{0} & \check{C}^{1} & \ldots & \check{C}^{p} & \check{C}^{p+1} & \check{C}^{p+2}
\end{array}.\label{QFT TTT}
\end{equation}
The action of the exterior derivative $d$ moves us one step up in the table (with two steps up always yielding zero because $d^2=0$), and the action of the \v Cech coboundary operator $\delta$ moves us one step to the right (with two steps right always yielding zero because $\delta^2=0$). Conversely, if an element lies beneath a zero entry (which means the locally-defined forms are closed), we can use the Poincar\'e lemma to move one step down,\footnote{We note that the result of doing so is not unique.} and if an element lies to the left of a zero entry, the existence of a partition of unity enables us to ``undo" $\delta$ and move one step to the left, analogous to the Poincar\'e lemma for ``undoing" $d$.\footnote{A partition of unity is a collection of functions $\{p_{\alpha}\}$ on the open sets such that $p_{\alpha}\geq 0$, $\sum_{\alpha}p_{\alpha}=1$, and $p_{\alpha}$ has compact support in $U_{\alpha}$. If $\{T_{\alpha_0\alpha_1\dots\alpha_r}\}\in \check{C}^r(\mathcal{U},\Lambda^q)$, then the object $\{S_{\alpha_0\alpha_1\dots\alpha_{r-1}}\}\in \check{C}^{r-1}(\mathcal{U},\Lambda^q)$ defined by $\{S_{\alpha_0\alpha_1\dots\alpha_{r-1}}\}=\{\Sigma_{\gamma} T_{\alpha_0\alpha_1\dots\alpha_{r-1}\gamma}p_{\gamma}\}$ satisfies $\{T\}=\delta\{S\}$.} 
The element in the bottom right of the tic-tac-toe table, which we denote by $\{K\}:=\delta\{A^0\}$, is a set of 0-forms defined on $(p+2)$-fold intersections, which is both $d$ and $\delta$ closed.  The importance of this object shall become clear after we have written down the action (phase), and so we postpone further discussion for now. 

The action shall be a sum of integrals of all of these locally-defined forms. We now describe how to obtain the set of chains on which to integrate these forms. Having applied $\mathrm{Sd}^n$, we thus far have chosen a set of $\mathcal{U}$-small $p$-chains $\{c_{p,\alpha}\}$ on which to integrate $\{A^p_{\alpha}\}$ (wherever a $\mathcal{U}$-small simplex lies in a double intersection, say $U_{\alpha\beta}$, simply make a choice to include this simplex in either $c_{p,\alpha}$ or $c_{p,\beta}$). Given each $c_{p,\alpha}$, its boundary can be written as a sum over $(p-1)$-chains which are contained in the double intersections of $U_{\alpha}$ with each of the other open sets, {\em viz.} $\partial c_{p,\alpha}=\sum_{\beta} c_{(p-1),\alpha\beta}$. 
By taking the boundary of each $c_{p,\alpha}$ and collecting terms defined on each double intersection, we thus obtain a set of $(p-1)$-chains $\{c_{(p-1),\alpha\beta}\}$, which are $\mathcal{U}$-small in the sense of being contained wholly in {\em double} intersections, on which we can integrate the local $(p-1)$-forms $\{A^{p-1}_{\alpha\beta}\}$.

Proceeding, given a set $\{c_{(p-q+1),\alpha_0\dots \alpha_{q-1}}\}$ of $(p-q+1)$-chains defined on $q$-fold intersections, we construct the appropriate $(p-q)$ chains in the obvious way: simply take the boundary, expressed as a sum of $(p-q)$-chains lying wholly in the $(q+1)$-fold intersections of our good cover. Thus, in analogy to how we started from the global $(p+1)$-form $\omega$ and constructed a tower of local forms, we can start from a $p$-cycle and construct a tower of $\mathcal{U}$-small chains of degree $p,p-1,\dots,0$, right down to a set of points (0-chains) $\{c_{0,\alpha_{0}...\alpha_{p}}\}$ defined on $(p+1)$-fold intersections.

We have now constructed all the objects that we need to write down a consistent action.
We define the action to be the following sum of integrals, of the locally-defined forms on the corresponding $\mathcal{U}$-small chains:
\begin{equation}
S[z]=\sum_{\alpha}\int_{c_{p,\alpha}}A^p_{\alpha}-\sum_{\alpha \beta}\int_{c_{(p-1),\alpha\beta}}A^{p-1}_{\alpha\beta}+\dots
+(-)^{p}\sum_{\alpha_{0}...\alpha_{p+1}}A^0_{\alpha_{0}...\alpha_{p}}(c_{0,\alpha_{0}...\alpha_{p}}).\label{WYA}
\end{equation}
One can show that this action is free of any ambiguities in degree $>0$, which potentially arise when there is a choice of local forms to integrate on a particular simplex. The argument is a rather technical digression, which we therefore reserve for Appendix \ref{ambiguities}. The essential idea behind this argument is that any ambiguity in forms of a given degree is removed by the presence of forms constructed in one degree lower, by virture of the relations coded in the tic-tac-toe table \ref{QFT TTT}.

However, once we get all the way down to the ambiguity in the 0-forms, it is no longer possible to remove the ambiguity by adding forms of one lower degree, since no such forms exist. Thus, there is a seemingly irremovable ambiguity in the presence of non-vanishing $(p+2)$-fold intersections, since then we can choose to evaluate one of $(p+2)$ different 0-forms on a 0-chain contained therein. This $0$-form ambiguity between different choices can, in general, be written as
\begin{equation}
S'-S=K_{\alpha_{0}...\alpha_{p+1}},
\end{equation}
where $K_{\alpha_{0}...\alpha_{p+1}}$ is an element of the \v Cech $(p+1)$-cochain $\{K\}:=\delta\{A^0\}$. For example, in $p=1$, the ambiguity occurs when a 1-simplex $\sigma$ is contained in a triple intersection, say $U_{\alpha\beta\gamma}$. In this case, choosing to integrate either $A_{\alpha}$, $A_{\beta}$, or $A_{\gamma}$ over the simplex $\sigma$ leads to actions that differ by $S'-S=A_{\alpha\beta}+A_{\beta\gamma}+A_{\gamma\alpha}$. Sure enough, the RHS is an element of $\{K\}=\delta\{A^0\}$.

Thus, consistency appears to require that $\{K\}$, which are, {\em a priori}, real-valued functions on $U_{\alpha_0\dots\alpha_{p+1}}$, must vanish.
In fact this is too strong, because only the action phase needs to be well-defined, so that each $K_{\alpha_{0}...\alpha_{p+1}}$ need only equal an integer. Even this seems to require a miracle, but it is, very nearly, a {\em fait accompli}. To see this, recall from the tic-tac-toe table (\ref{QFT TTT}) that $\{K\}$ is closed under both $d$ and $\delta$. Given $\{K\}$ is valued in 0-forms, $d$-closure implies that each $K_{\alpha_{0}...\alpha_{p+1}}$ is, in fact, constant on the $(p+2)$-fold intersection on which it is defined. Thus, $\{K\}$ defines an element in $\check{C}^{p+1} (\mathcal{U},\mathbb{R}) \subset \check{C}^{p+1} (\mathcal{U},\Lambda^0)$. So the only miracle that need occur is that the real constants $\{K\}$ be integers. Moreover, $\delta$-closure implies $\{K\}$ is a \v Cech $(p+1)$-co\textit{cycle}, and thus defines a cohomology class, $[\{c\}]\in \check{H}^{p+1}(G/H,\mathbb{R})$.
Therefore, in the usual language of cohomology, $[\{c\}]\in \check{H}^{p+1}(G/H,\mathbb{Z})$ must be an integral class for the action to define a well-defined path integral ~\cite{Alvarez:1984es} for all p-cycles in $G/H$.

The desired integrality of the \v Cech $(p+1)$-cocycle $[\{c\}]$ is equivalent, via the \v Cech-de Rham isomorphism, to the requirement that $\omega$ be an integral $(p+1)$-form, 
\begin{equation}
[\omega]\in H^{p+1}(G/H,\mathbb{Z}).
\end{equation}
The \v Cech-de Rham isomorphism can be seen from the tic-tac-toe table (\ref{QFT TTT}). The external row of the tic-tac-toe table is reserved for real-valued \v Cech cocycles, such as $\{K\}$, on which $\delta$ has non-vanishing cohomology (since using the partition of unity construction would take us out of the space of constant functions). The external column is reserved for globally-defined forms, on which $d$ has non-vanishing cohomology (since using the Poincar\'e lemma would us out of the space of globally-defined forms). The tic-tac-toe table allows us to move between \v Cech and de Rham cocycles of the same degree, by successive applications of (say) the exterior derivative $d$ and a partition of unity to undo $\delta$ (if going from \v Cech to de Rham, that is, from bottom right to top left of the table). This
provides an explicit construction of the isomorphism on cohomology ~\cite{bott1995differential}. 

For a more physical way to see the integrality condition, consider the case $p=1$.  In $p=1$, the action specializes to that first introduced by Wu \& Yang ~\cite{Wu:1976qk} in their formulation of the action for the Dirac monopole, with the additional terms in the action due to the 0-forms $\{A^0\}$ being precisely the transition function insertions that Wu \& Yang introduced. In this simplest case, the integral of $\omega$ over a 2-cycle $y$ in $G/H$, which can be written in terms of a sum over $\mathcal{U}$-small 2-chains $c_{2,\alpha}$ contained in $U_{\alpha}$, viz. $\int_y \omega = \sum_{\alpha} \int_{c_{2,\alpha}} \omega|_{\alpha}=\int_{c_{2,\alpha}} dA^1_{\alpha}$. Using Stokes' theorem twice, one obtains
\begin{equation}
\int_y \omega = \sum_{\alpha\beta\gamma} K_{\alpha\beta\gamma}, \label{1d flux}
\end{equation}
where on the RHS we sum (the appropriate number of times) over those triple intersections which have non-vanishing intersection with the image of $y$. Since the RHS is an integer for any 2-cycle, $\omega$ is therefore an integral 2-form. This generalizes to higher $p$.

It is instructive, at this point, to return to our earlier, prototypical examples. We now see that, whenever $\omega$ lies in a non-trivial de Rham cohomology class, we obtain a ``quantization condition" on the coupling in the action, just as we did for the Dirac monopole. But if $\omega$ is de Rham exact, as in the case of the Landau problem, the integral over any $(p+1)$-cycle vanishes automatically, and there will 
be no quantization condition on the coupling in the action.
Moreover, the \v Cech-de Rham isomorphism guarantees that the correspondence goes the other way, such that for every integral $(p+1)$-form, there exists a corresponding choice of integer \v Cech $(p+1)$-cocycle, and hence a well-defined action phase.

How does the definition (\ref{WYA}) of the action for a WZ term connect with Witten's construction, in the special cases where the latter may be used?
When the action (\ref{WYA}) is evaluated for a cycle $z=\partial b$ that is in fact the boundary of a $(p+1)$-chain $b$, one can show that the action phase does indeed reduce to Witten's $\exp (2\pi i \int_{b} \omega)$. To see this, consider to begin with the special case $p=1$, and consider a 1-cycle $z$ whose image intersects three double intersections $U_{\alpha\beta}$, $U_{\beta\gamma}$, and $U_{\gamma\alpha}$. The action (\ref{WYA}) may in this case be written as
\begin{equation}
S[z] = \int_{c_{1,\alpha}}A^1_{\alpha}-A^0_{\alpha\beta}(c_{0,\alpha\beta})+\int_{c_{1,\beta}}A^1_{\beta}-A^0_{\beta\gamma}(c_{0,\beta\gamma})+\int_{c_{1,\gamma}}A^1_{\gamma} -A^0_{\gamma\alpha}(c_{0,\gamma\alpha}). \label{WY action on cycle}
\end{equation}
If $z$ is in fact the boundary of a 2-chain $b$, then the three open sets share a non-vanishing triple intersection, $U_{\alpha\beta\gamma}\neq \varnothing$. Let $\tilde{c}_{0}$ be an arbitrary 0-chain (point) whose image is contained in this triple intersection, $\mathrm{Im}\ \tilde{c}_{0} \in U_{\alpha\beta\gamma}$. After subdivision, the 2-chain $b$ may be written as the sum of three $\mathcal{U}$-small 2-chains, $b=b_{\alpha}+b_{\beta}+b_{\gamma}$ each contained within the open sets, with common point $\tilde{c}_{0}$ in the triple intersection. Moreover, the boundaries of these 2-chains define a set of 1-cycles, $z_{\alpha}=\partial b_{\alpha}$ etc., which are necessarily also $\mathcal{U}$-small.

One can then show, using $\delta\{A^1\}=\{dA^0\}$, that (\ref{WY action on cycle}) is equal to
\begin{equation}
S[z]=\int_{z_\alpha} A^1_{\alpha} + \int_{z_\beta} A^1_{\beta} + \int_{z_\gamma} A^1_{\gamma} - A^0_{\alpha\beta}(\tilde{c})-A^0_{\beta\gamma}(\tilde{c})-A^0_{\gamma\alpha}(\tilde{c}).
\end{equation}
Using Stokes' theorem on each open set, this reduces to
\begin{equation}
S[z]=\int_{b_\alpha} \omega_{\alpha} + \int_{b_\beta} \omega_{\beta} + \int_{b_\gamma} \omega_{\gamma} - K_{\alpha\beta\gamma}(\tilde{c}),
\end{equation}
where we have also used $\{K\}=\delta\{A^0\}$. But since the 2-form $\omega$ is globally-defined, and $K_{\alpha\beta\gamma}$ is constant throughout triple intersections, we have simply that
\begin{equation}
S[z]=\int_b \omega - K_{\alpha\beta\gamma}. 
\end{equation}
Moreover, we know that the \v Cech 2-cocycle $\{K\}$ must be valued in integers, for consistency's sake. Hence, 
\begin{equation}
e^{2\pi i S[z]}=e^{2\pi i \int_b \omega}, \quad z=\partial b. \label{witten}
\end{equation}
That is, the action phase prescribed by (\ref{WYA}) does indeed reduce to that prescribed by the Witten construction. This argument generalizes straightforwardly to generic boundaries and higher dimensions.

The action (\ref{WYA}) we have defined, which is free of ambiguities over which local forms to integrate, is moreover well defined on the fundamental class, ergo is a well-defined $\mathcal{O}$-invariant of $\Sigma^p$. Indeed, let $z$ and $z^\prime$ be two fundamental cycles on $\Sigma$, and let $\phi_* z$ and $\phi_* z^\prime$ be the corresponding cycles on $G/H$. Suppose that $n$ subdivisions are enough to split the simplices in both $z$ and $z^\prime$ sufficiently.\footnote{Once the chains are $\mathcal{U}$-small, further applications of the subdivision operator do not change the value of the action.}
The difference in the action for the two cycles is then $\delta S = S[\mathrm{Sd}^n \phi_* z] - S[\mathrm{Sd}^n \phi_* z^\prime]$.

Using the facts that all maps are homomorphisms, that the 2 cycles $z$ and $z^\prime$ are homologous, and that $\partial$ is a natural map, this simplifies to
\begin{gather}
\delta S = S [ \mathrm{Sd}^n \phi_* (z-z^\prime)] = S[\mathrm{Sd}^n \phi_* \partial b] = S[\partial \mathrm{Sd}^n \phi_*  b].
\end{gather}
The shift in the action is thus expressed as a contribution on a boundary, which we have already shown (\ref{witten}) reduces to
$\int_{\mathrm{Sd}^n \phi_*  b} \omega = \int_{\phi_* \mathrm{Sd}^n b} \omega = \int_{ \mathrm{Sd}^n b} \phi^* \omega$, where we used the fact that the subdivision operator is also a natural map.
Now, $\mathrm{Sd}^n b$ is a $(p+1)$-chain on $\Sigma^p$, so pulling back $\omega$ to the constituent simplices and integrating must yield 0.
Hence the action $S$ is well-defined on $[\Sigma^p]$ and $\Sigma^p$.

Before we continue, let us pause to give more detail on the interpretations on the mathematics and physics sides in $p=1$.
Mathematically, 
given the quantization condition on the closed 2-form $\omega$, the collection of 1-forms $\{A^1_{\alpha}\}$ defines a connection on a $U(1)$ principal fibre bundle with base $G/H$, with $\omega$ (or rather its pullback via the bundle map) being the curvature of that connection \cite{10.1007/BFb0092438}.\footnote{In higher dimensions  $p>1$, $A_{\alpha}$ is a $p$-form generalization of a background gauge field, also known as a $p$-form connection ~\cite{Freed:2006yc}. The action phase becomes the appropriate higher dimensional generalization of the holonomy ~\cite{2013arXiv1303.6457B}.} The quantization of $\omega$ corresponds to the condition, necessary for the existence of the bundle,  that the first Chern class $c_1$ be an integer. The action phase we have defined using local forms on $G/H$ is, from the bundle perspective, nothing but the holonomy of the connection on the cycle $z$.
In physical terms, $\{A^1_{\alpha}\}$ constitutes a $U(1)$ gauge field, with $\{A^0_{\alpha \beta}\}$ denoting gauge transformations on the overlaps, and $\omega$ being the gauge invariant electromagnetic field strength. The integrality of $\omega$ means that the magnetic flux through any 2-cycle is quantized. When $G/H=SO(3)/SO(2)$ is homeomorphic to the 2-sphere, such that the cohomology is generated by a single class, this is equivalent to Dirac's quantization condition on the charge of a magnetic monopole.

\subsection{Invariance and the Manton condition \label{invariance}
} 
With a consistent action for WZ terms in hand, we may now turn to the issue of invariance under the action of the Lie group $G$.
Indeed, in quantum field theory we would like the $G$-action on $G/H$ to be a symmetry of the path integral. We can use the left Haar measure to define the path integral measure and so (at least in the absence of fermions and associated anomalies) we can, in what follows, concentrate our attention on $G$-invariance of the action phase.

We already argued in \S \ref{consistency} that,
when the worldvolume cycle $z$ is the boundary of a $(p+1)$-chain, $z=\partial b$, the action can be written as the integral of a $(p+1)$-form $\omega$ over $b$, and so is invariant under the $G$-action when $\omega$ is invariant under pullback by the action of $G$. (As usual, we call such a form a {\em $G$-invariant} form on $G/H$.)
However, when the worldvolume cycle is homologically non-trivial, the action must be written in the form of equation \ref{WYA}, with contributions from a slew of locally-defined $p,p-1,\dots$ forms, so $G$-invariance of the action does not necessarily follow from $G$-invariance of $\omega$ alone. What is worse, it is difficult, {\em a priori}, to even imagine how a simple condition for $G$-invariance can be obtained, given that the pullback of forms by the action of $G$ on $G/H$ takes locally-defined forms out of the patches on which they are defined. Thus, there is no simple notion of $G$-action on, let alone $G$-invariance of, locally-defined forms. Nonetheless, there is a well-defined action of the Lie algebra of $G$ on locally-defined forms, given by the Lie derivative. By requiring invariance of (\ref{WYA}) under this infinitesimal action, we will be able to obtain a necessary and sufficient condition for invariance when $G$ is connected.

Let us start by considering, for simplicity, 
the variation of the action (\ref{WYA}) when $p=1$ under an infinitesimal $G$ transformation, generated by vector field $X$ on $G/H$. A 1-cycle that is not the boundary of a 2-chain in $G/H$ must intersect at least three double intersections in a good cover of $G/H$, so let us consider this minimal non-trivial possibility.
The action for a cycle $z$ which intersects three double intersections $U_{\alpha\beta}$, $U_{\beta\gamma}$, and $U_{\gamma\alpha}$ is given by (\ref{WY action on cycle}), except that the triple intersection is now taken to vanish, $U_{\alpha\beta\gamma}=\varnothing$.
The infinitesimal variation of the action is
\begin{equation}
\delta_X S[z] = \int_{c_{1,\alpha}}L_{X}A^1_{\alpha}-L_{X}A^0_{\alpha\beta}(c_{0,\alpha\beta})+\int_{c_{1,\beta}}L_{X}A^1_{\beta}-L_{X}A^0_{\beta\gamma}(c_{0,\beta\gamma})+\int_{c_{1,\gamma}}L_{X}A^1_{\gamma} -L_{X}A^0_{\gamma\alpha}(c_{0,\gamma\alpha}),
\end{equation}
where $L_X$ is the Lie derivative. Applying Cartan's formula $L_X=d\iota_X+\iota_Xd$ to the local forms appearing in the action, we have
\begin{equation}
L_{X}A^1_{\alpha}=\iota_{X}\omega_{\alpha}+d\iota_{X}A^1_{\alpha},\quad \mathrm{and} \quad L_{X}A^0_{\alpha\beta}=\iota_X dA^0_{\alpha\beta}=\iota_{X}(A^1_{\alpha}-A^1_{\beta}),
\end{equation}
since $A^0_{\alpha\beta}$, \&{\em c.}, are just 0-forms in $p=1$, and since $\delta\{A^1\}=\{dA^0\}$. Hence, integrating and using Stokes' theorem, we are left with
\begin{equation}
\delta_X S[z] = \int_{c_{1,\alpha}}\iota_{X}\omega_{\alpha}+\int_{c_{1,\beta}}\iota_{X}\omega_{\beta}+\int_{c_{1,\gamma}}\iota_{X}\omega_{\gamma}=\int_z \iota_{X}\omega,
\end{equation}
where in the second step we have used the fact that $\omega$, and therefore $\iota_X \omega$, is globally-defined.
By a straightforward generalization, this applies for any 1-cycle $z$ in $G/H$.

This argument for $p=1$ generalizes straightforwardly to $p>1$. For example, in $p=2$, a consistent topological term corresponds to a global closed 3-form $\omega$ such that $\omega_{\alpha}=dA^2_{\alpha}$ on patches, for locally-defined 2-forms  $\{A^2_{\alpha}\}$. On double intersections we have
$A^2_{\alpha}-A^2_{\beta}=dA^1_{\alpha\beta}$ for locally-defined 1-forms $\{A^1_{\alpha\beta}\}$, which in turn satisfy $A^1_{\alpha\beta}+A^1_{\beta\gamma}+A^1_{\gamma\alpha}=dA^0_{\alpha\beta\gamma}$ on triple intersections for 0-forms $\{A^0_{\alpha\beta\gamma}\}$.
Consider, for simplicity, a 2-cycle $z$ contained within four open sets $U_{\alpha}$, $U_{\beta}$, $U_{\gamma}$ and $U_{\delta}$,\footnote{This is the minimal possibility for a non-trivial cycle in $p=2$.} which we write as a sum of $\mathcal{U}$-small 2-chains, $z=c_{2,\alpha}+c_{2,\beta}+c_{2,\gamma}+c_{2,\delta}$. The boundaries of these 2-chains provide the 1-chains over which we will integrate $\{A^1_{\alpha\beta}\}$ 
(for example, $\partial c_{2,\alpha}$ is a sum of $\mathcal{U}$-small 1-chains contributing to $c_{1,\alpha\beta}$, $c_{1,\alpha\gamma}$, and $c_{1,\alpha\delta}$, \&{\em c.}), and the boundaries of the resulting 1-chains provide the points on which we evaluate $\{A^0_{\alpha\beta\gamma}\}$.
The action (\ref{WYA}) for this cycle is then
\begin{equation}
S[z]=\sum_{\alpha}\int_{c_{2,\alpha}}A^2_{\alpha}-\sum_{\alpha\beta}\int_{c_{1,\alpha\beta}}A^1_{\alpha\beta}+\sum_{\alpha\beta\gamma}A^0_{\alpha\beta\gamma}(c_{0,\alpha\beta\gamma}),
\end{equation}
where we sum over all 2-chains, 1-chains, and 0-chains just described.
Taking the Lie derivatives, using relations (from the tic-tac-toe table (\ref{QFT TTT})) such as $\delta\{A^2\}=\{dA^1\}$, and using Cartan's formula, we obtain
\begin{equation}
L_{X}A^2_{\alpha}=\iota_{X}\omega_{\alpha}+d\iota_{X}A^2_{\alpha}, \quad L_{X}A^1_{\alpha\beta}=\iota_{X}(A^2_{\alpha}-A^2_{\beta})+d\iota_{X}A^1_{\alpha\beta}, \quad L_{X}A^0_{\alpha\beta\gamma}=\iota_{X}\left(A^1_{\alpha\beta}+A^1_{\beta\gamma}+A^1_{\gamma\alpha}\right).
\end{equation}
Again using Stokes' theorem, the variation of the action reduces to
\begin{equation}
\delta_X S[z]=\int_{c_{2,\alpha}}\iota_{X}\omega_{\alpha}+\int_{c_{2,\beta}}\iota_{X}\omega_{\beta}+\int_{c_{2,\gamma}}\iota_{X}\omega_{\gamma}+\int_{c_{2,\delta}}\iota_{X}\omega_{\delta} =\int_{z} \iota_{X}\omega,
\end{equation}
exactly as we found for $p=1$. The equations relating the \v Cech-de Rham double cochains which appear in our action (which follow from consistency) will guarantee similar cancellations in general $p$, such that 
\begin{equation}
\delta_X S[z]=\int_{z} \iota_{X}\omega \label{variation of action}
\end{equation}
holds in general $p$.\footnote{We note that the shift in the action reduces to an AB term. In fact this remains true even when $G$ is disconnected.}

For the action to be invariant under all infinitesimal $G$ transformations, $\int_z \iota_{X}\omega$ must therefore vanish for all vector fields $X$ that generate the $G$-action, on all $p$-cycles $z\in Z_p(M,\mathbb{Z})$. From (\ref{dR}) we conclude that $\iota_{X}\omega$ must be an exact form for all such $X$. In other words, the interior product of $\omega$ with each vector field must lie in the trivial de Rham cohomology class
\begin{equation}
[\iota_{X}\omega]_{dR}=0, \quad \forall X, \label{Manton condition}
\end{equation}
where $[\cdot]_{dR}$ indicates the de Rham cohomology class of a form. 

We call the condition (\ref{Manton condition}) the `Manton condition', since its failure in the case of $p=1$ and $G/H\cong T^2$, which corresponds to the quantum mechanics of a particle on the torus in a uniform $B$ field,
leads to the breaking of translation invariance, an `anomaly' that was first appreciated by Manton ~\cite{Manton:1983mq,Manton:1985jm}. Manton's derivation relied on an explicit solution for the wavefunctions of the corresponding quantum mechanics problem. We now see that it has a rather broad generalization to any homogeneous space sigma model in quantum field theory, which can be phrased in terms of a simple, geometric condition, whose derivation, serendipitously, does not require a solution of the field theory, but can be derived directly from the topological action. Explicitly, it may be understood as arising from the requirement that the action be invariant for \textit{all} cycles. This is non-trivial in a general quantum field theory, because even defining the action for all cycles is, as we have seen, non-trivial.

At the beginning of this Subsection, we saw that, for homologically trivial cycles, $G$-invariance of the action follows from $G$-invariance of the $(p+1)$-form $\omega$. At the infinitesimal level, this is equivalent to the vanishing of the Lie derivatives, $L_X\omega=0$. How does this relate to the Manton condition?
Applying Cartan's formula to $\omega$, which is closed, tells us that $L_X \omega = d \iota_X \omega$, and so left-invariance of $\omega$ only implies that $i_X\omega$ is closed, but not necessarily that it is exact.
Hence, the Manton condition is a \textit{stronger} condition than $G$-invariance of $\omega$. When there exist non-trivial homology cycles, such that $H_d(G/H,\mathbb{Z})\neq 0$, the weaker condition $L_X\omega=0$ is insufficient for the existence of a $G$-invariant WZ term, as we have already seen from the torus example. At least at the infinitesimal level, the Manton condition strengthens the necessary condition of vanishing Lie derivatives to a necessary and sufficient condition.

As a consistency check, we show that 
when all cycles are boundaries, {\em i.e.} when $H_d(G/H,\mathbb{Z})=0$ (such that the Witten construction can be applied), the Manton condition is equivalent to $L_X\omega = 0$. We have already shown that the action (\ref{WYA}) can in this case be written as $S[z]=\int_{b} \omega$, where $b$ is any $(p+1)$-chain such that $z=\partial b$.
The variation of the action is then $\delta_X S[z]=\int_b L_X\omega$, which must vanish on all chains, so invariance is obtained if and only if $L_X\omega=0$, using (\ref{dR}). 
To show that (in this situation) this is equivalent to the Manton condition, we note firstly that
if $H_p(G/H,\mathbb{Z})=0$, then $H_p(G/H,\mathbb{R})=0$ too. But the real (smooth singular) cohomology is simply the dual of real homology, and is moreover isomorphic to the $p$th de Rham cohomology. The latter therefore vanishes, 
and hence the closed $p$-form $\iota_X\omega$ is automatically exact.
Therefore, our procedure is seen to be equivalent to (the homological version of) Witten's construction in all cases where the latter is valid.

We have shown that the Manton condition is necessary and sufficient for invariance under infinitesimal $G$ transformations generated by vector fields $X$. By arguments similar to those given in \S \ref{ab}, this invariance extends at the group level both to the image of the exponential map in $G$ and thence to the component connected to the identity.
The Manton condition is thus both necessary and sufficient at the group level when $G$ is connected, as we here assume.

\subsection{Injectivity of WZ terms \label{injectivity}} 

We have shown that there exists a consistent, $G$-invariant topological term for every closed, integral $(p+1)$-form $\omega$ on $G/H$ satisfying the Manton condition. We should be careful, however, that this topological term may not be unique. Indeed, $\omega=0$ is such a form, and corresponds to, in general, a whole set of topological terms, namely the AB terms of \S \ref{ab}. Thus, any non-uniqueness in $\omega$ corresponds to an AB term. And so, there is a map from such $(p+1)$-forms $\omega$ to WZ terms. We now show that this map is injective.

To do so, let $b$ be any $(p+1)$-chain in $G/H$. Two $(p+1)$-forms $\omega$ and $\omega^\prime$ could yield the same action phase only if they agree on cycles $z = \partial b$ for all $b$. But for such $p$-cycles which are the boundaries of $(p+1)$-chains, we can write the action phase directly using the Witten construction as $\exp{2\pi i \int_{b} (\omega^\prime - \omega)}$, whence we would need
$\int_{b} (\omega^\prime - \omega) \in \mathbb{Z}$ in order for the two action phases to coincide.
In fact, we will now show that $\int_b (\omega^\prime - \omega)$ would have to vanish  for all $(p+1)$-chains $b$. To wit, given any $(p+1)$-simplex $\sigma: \Delta^{p+1} \rightarrow G/H$,
consider the maps $T_t: \Delta^{p+1} \rightarrow \Delta^{p+1} : x
\mapsto tx$, with $t \in [0,1]$ and form the simplex
$\sigma_t = \sigma \circ T_t$. The simplex $\sigma_t$ defines a chain, so the integral $\int_{\sigma_t}  (\omega^\prime - \omega) $
must be a continuous, integer-valued function on $t\in [0,1]$. But
$\int_{\sigma_0}  (\omega^\prime - \omega) = 0$. Therefore, by continuity, $0=\int_{\sigma_1}  (\omega^\prime - \omega) =
\int_{\sigma}  (\omega^\prime - \omega)$. The integral thus vanishes on all 
simplices and thence vanishes on all chains. By (\ref{dR}), this means
that $ \omega^\prime = \omega $.
In other words, the only topological terms which can lead to the same action phase on all cycles have $\omega=0$, {\em i.e.} they are of AB type (where we know from \S \ref{ab} that the injectivity requirement leads to a quotient by closed integral $d$ forms).

\subsection{The classical limit and Noether currents \label{noether currents}} 

By Noether's theorem, the invariance of the action under the action of a Lie algebra $\mathfrak{g}$ implies the existence of conserved currents, at least at the classical level. We now explore the status of these currents in the presence of topological terms. We find an interesting connection with the Manton condition. To wit, whilst the weaker condition of $G$-invariance of $\omega$ ensures $G$-invariance of the equations of motion, a corresponding Noether current exists only when the stronger Manton condition is satisfied. Thus, the Manton condition, which we derived as the condition for $G$-invariance of the quantum theory, has a physical vestige even in the classical limit.

To derive the Noether currents associated with
$G$-invariance, we take the variation of the action (\ref{WYA}) induced by the vector field $\epsilon^{a}\left(x\right)X_{a}$, where $a$ runs over the vector fields generating $G$, for some non-constant functions $\epsilon^{a}\left(x\right)$. Recall that when the $\epsilon^a$ are constants the action is $G$-invariant, and so the variation of the action will be proportional to the 1-form $d\epsilon^{a}$.
We can then read off the Noether current and deduce that it is conserved on the classical equations on motion.

We first consider, for simplicity, $p=1$, with the action given by (\ref{WY action on cycle}) ({\em i.e.} evaluated on a 1-cycle intersecting three double intersections). The variation in the local 1-forms is given by $L_{\epsilon^a {X_a}}A^1_{\alpha}=\epsilon^a L_{X_a} A^1_{\alpha}+\iota_{X_a} A^1_{\alpha}d\epsilon^a $. Since the action is $G$-invariant, the Manton condition holds and there exists a set of globally-defined 0-forms, $f_a$, one for each vector field $X_a$, such that 
\begin{equation}
\iota_{X_a} \omega=df_a \label{definition f}
\end{equation}
Therefore, we may write 
\begin{equation}
L_{\epsilon^a {X_a}}A^1_{\alpha}=d[\epsilon^a (f_a+\iota_{X_a} A^1_{\alpha})]- f_{a}d\epsilon^{a}. 
\end{equation}
The variation in the 0-forms that appear in the action is
\begin{equation}
L_{\epsilon^a {X_a}}A^0_{\alpha\beta}=\epsilon^a \iota_{X_a} (A^1_{\alpha}-A^1_{\beta}).
\end{equation}
The only piece that survives in the variation of the action (noting that we have already used the fact that $X$ generates a symmetry by writing $\iota_{X_a} \omega=df_a$) is
\begin{equation}
\delta_{\epsilon X}S[z]=-\int_z f_a d\epsilon^a.
\end{equation}
When the equations of motion hold, any field variation vanishes. We can integrate by parts to deduce 
that $df_a=0$, and so the functions $Q_a=f_a$ are conserved on-shell and may be identified as the 0-form Noether charges in $p=1$.

In general $p$, the $f_a$ are $(p-1)$-forms, and the variation of the action (on a generic $p$-cycle $z$) induced by $\epsilon^{a}\left(x\right)X_{a}$ is
\begin{equation}
\delta_{\epsilon X}S[z]=-\int_{z} d\epsilon^a \wedge f_a.
\end{equation}
Again, when the equations of motion hold, we deduce that
\begin{equation}
df_a=0, \label{current conservation}
\end{equation}
and so may identify the $f_a$ as the $(p-1)$-form Noether currents corresponding to $G$-invariance, with  (\ref{current conservation}) being the equations for current conservation on-shell.\footnote{It is usual, in the presence of a metric, to define a Noether current as a 1-form via the Hodge dual, {\em viz.} $j_a=\star f_a$, with $df_a=0$ being equivalent to $\mathrm{div} j_a=0$. But since a metric is not presumed to be available, we prefer to formulate Noether's theorem directly in terms of the $(p-1)$-forms $f_a$.} In a Lorentzian theory, we may obtain the conserved Noether charges by integrating the $(p-1)$ forms $f_a$ over spatial hypersurfaces.

Thus, $G$-invariant topological terms in the action result in a shift of the conserved currents. Now, let us examine more closely the r\^{o}le played by the Manton condition in the argument just given. 
If the Manton condition does not hold, then the $(p-1)$-forms $f_a$, while guaranteed to exist locally by the Poincar\'e lemma, are not globally-defined. Thus, there is no way to patch together the locally-defined currents to make a {\em bona fide}, globally-defined, conserved current. 

Nevertheless, it is still possible that classical dynamics is $G$-invariant, even when the Manton condition fails to hold. Indeed, we know that the equations of motion only feature the $(p+1)$-form $\omega$. Thus, the classical dynamics will be invariant under the weaker condition of $G$-invariance of $\omega$, but there will be no conserved current associated to $X$ unless $\iota_X \omega$ is also exact.

For the example of particle motion on the torus in the presence of a uniform magnetic field, specified by the electromagnetic field strength $Bdx\wedge dy$ (which is invariant under $U(1)\times U(1)$), with $x\sim x+1$ and $y\sim y+1$, the equations of motion have $U(1)\times U(1)$ symmetry, but there are no conserved currents even classically, and in the quantum theory the true symmetry is at most a discrete subgroup of $U(1)\times U(1)$.
Indeed, as we saw in \S \ref{QM examples}, the action phase for a cycle wrapping the $y$ direction is $e^{2\pi i B x_0}$, which is invariant under a translation $x\rightarrow x+a$ only if $a \in \{0,1/B, 2/B, \dots, (B-1)/B\} \cong \mathbb{Z}/B\mathbb{Z}$ (we recall that consistency forces $B$ to be an integer). A similar argument for a cycle wrapping the $x$ direction shows that the full unbroken subgroup is $(\mathbb{Z}/B\mathbb{Z})^2$. The order $B^2$ of this subgroup is of course the degeneracy of the ground state Landau level in the presence of a uniform $B$ field with periodic boundary conditions.

Finally, it is interesting to note from (\ref{definition f}) that the contribution from the topological term to the current is conserved \textit{off}-shell if $\iota_X \omega=0, \ \forall X$. This can only happen for AB terms, as the following argument shows. 
Since $G$ acts transitively on $M= G/H$, the vector fields $\{X\}$ span the tangent space $TM$ at each point. Moreover, $\iota_X (\omega) = 0$ implies $\iota_{X_1}(\iota_{X_2} ... (\iota_{X_d} (\omega))) = 0$, where $X_1,X_2,\dots,X_d\in TM$. Hence, $\omega$ is a $(p+1)$-fold skew-symmetric linear map that yields zero on all elements of $T^{p+1} M$, that is, it is the zero map. Thus, off-shell current conservation implies (and is implied by) $\omega = 0$, such that the off-shell conserved currents are in one-to-one correspondence with the AB terms classified in \S \ref{ab}. In retrospect this is hardly surprising, since an arbitrary infinitesimal variation of fields takes cycles into homologous cycles, on which the value of an AB term (but not a WZ term) is unchanged.

\subsection{Examples}
\subsubsection{The chiral lagrangian \label{chiral}} 
In the case of the chiral lagrangian, describing the low energy limit of QCD with 3 massless flavours of quarks, we have $p=4$ and $G/H =SU(3) \times SU(3)/SU(3) \cong SU(3)$. Since $H^4_{dR} (SU(3)) = 0$, the homological version of Witten's construction can be employed and our classification suggests that the topological terms correspond to closed, integral, $SU(3)\times SU(3)$-invariant 5-forms. Now, no such form can be the exterior derivative of an $SU(3)\times SU(3)$-invariant 4-form, because $SU(3) \times SU(3)/SU(3)$ is a symmetric space and  all invariant forms are closed on such a space \cite{schwarz1994}. Hence the topological terms correspond to integral classes in the Chevalley-Eilenberg cohomology of invariant forms \cite{0031.24803} ({\em cf.}~\S\ref{compute}), which in turn (since $G$ is compact and connected) correspond to integral classes in $H^5_{dR} (SU(3)) = \mathbb{Z}$. Thus, there is a WZ term given by integrating an integral $SU(3)\times SU(3)$-invariant 5-form over any 5-chain bounding the 4-cycle $\phi_* [\Sigma^4]$.

Though these results are superficially identical to those obtained by homotopical arguments by Witten \cite{Witten:1983tw}, there is a small, but significant, difference. If one fixes the worldvolume to be homeomorphic to $S^4$, then one may define the action by integrating the 5-form on a 5-disk and the possible ambiguity that arises from the choice of 5-disk may be removed from the action phase by insisting that the integral of the 5-form over any 5-sphere be integral. 
In our homological language, such 5-spheres correspond to a restricted set of 5-cycles; it turns out that a 5-form whose integral over {\em every} cycle is an integer has an integral over this restricted set of cycles given by an {\em even} integer. Thus, if one is only interested in topological terms in a theory with worldvolume $S^4$, one may safely take the 5-form to be a `half-integral form'.\footnote{In fact, the same is true for any worldvolume manifold admitting a spin structure \cite{Freed:2006mx}.} 

This fact, which corresponds to Witten's observation \cite{Witten:1983tw} that `the normalization of $\omega $ is a subtle mathematical problem', follows straightforwardly, provided one is willing to accept that $\pi_5 (S^3) = \pi_4 (S^3) = \mathbb{Z}/2\mathbb{Z}$. Since $SU(3)$ may be regarded as a fibre bundle $S^3 \cong SU(2) \rightarrow SU(3) \rightarrow SU(3)/SU(2) \cong S^5$,\footnote{To see this, note that $SU(3)$ has a transitive action on unit vectors in $\mathbb{C}^3$ with any such vector stabilized by  some $SU(2) \subset SU(3)$.} we have
a long exact sequence in homotopy, as well as a long exact sequence in homology arising from the Serre spectral sequence. 
Now, the Hurewicz map $h$ is a natural map from homotopy to homology, meaning that we have a commutative diagram
\begin{gather}
\begin{CD}
\pi_5 (S^3) @>>> \pi_5 (SU(3)) @>>> \pi_5 (S^5) @>>> \pi_4 (S^3) @>>> \pi_4 (SU(3))\\
@. @VhVV     @VhVV   @. @.     \\
H_5 (S^3) @>>> H_5 (SU(3)) @>>> H_5 (S^5)  @>>> H_4 (S^3) 
\end{CD}
\end{gather}
given explicitly by
\begin{gather}
\begin{CD}
\mathbb{Z}/2\mathbb{Z} @>>> \mathbb{Z}  @>>> \mathbb{Z}  @>>> \mathbb{Z}/2\mathbb{Z} @>>> 0 \\
@. @VhVV     @VhVV   @. @.     \\
0 @>>> \mathbb{Z} @>>> \mathbb{Z}  @>>> 0 
\end{CD}
\end{gather}
The right-hand arrow in the square is an isomorphism by the Hurewicz theorem, while the bottom arrow in the square is an isomorphism. A bit of algebraic {\em su doku} shows that the top arrow in the square can only be multiplication by 2, so the Hurewicz map $\pi_5 (SU(3)) \rightarrow H_5 (SU(3))$ must be given by multiplication by 2 as well. Hence the integral of the 5-form over a cycle corresponding to a 5-sphere results in an even integer.

\subsubsection{Beyond the minimal composite Higgs model \label{beyond minimal composite higgs}
} 
We now use our results to briefly discuss two WZ terms that have been claimed to exist in the literature on non-minimal composite Higgs models and exploited for phenomenological purposes ~\cite{Gripaios:2009pe,Gripaios:2016mmi}. 
Recall that we have already seen there \textit{is} a topological term for the minimal composite Higgs model, but that it is of AB type.

For our first example, consider the model based on the homogeneous space $G/H=SO(6)/SO(5)\cong S^5$.\footnote{The principal appeal of this model, compared to the minimal model, is that, since $SO(6)$ is locally isomorphic to $SU(4)$, one can easily imagine a UV completion in the form of a (technically natural) strongly coupled gauge theory with fermions.} Our classification  shows that the effective lagrangian contains a WZ term, corresponding to a closed, integral 5-form $\omega$, proportional to the $SO(6)$-invariant volume form on $S^5$, in agreement with ~\cite{Gripaios:2009pe}. The Manton condition is obviously satisfied, because the fourth de Rham cohomology of $S^5$ vanishes; for the same reason, there are no AB terms. Since the fourth homology vanishes, we can follow Witten's construction and write the action on $\Sigma \cong S^4$ as the (manifestly $SO(6)$-invariant) integral of $\omega$ over a 5-ball whose boundary is our worldvolume.

In the second example, our classification shows that there is in fact no WZ term. This example is based on homogeneous space $G/H=(SO(5)\times U(1))/ SO(4)\cong S^4\times S^1$ ~\cite{Gripaios:2016mmi}.\footnote{The principal appeal of this model was that the additional pseudo-Goldstone boson could explain a $\gamma\gamma$ resonance observed at $750$ GeV in LHC data. Sadly, this appears to have been a statistical fluctuation.} The problem here is that the closed, integral 5-form employed in ~\cite{Gripaios:2016mmi} (which, again, is just proportional to the volume form) does not satisfy the Manton condition. Indeed, the interior product of the volume form on $S^4\times S^1$ with the vector field generating the $U(1)$ factor can be pulled back to $S^4$ resulting in a form which is proportional to the volume form on $S^4$. This form is obviously not exact, and since the exterior derivative commutes with the pullback, the Manton condition cannot be satisfied.

The erroneous term in ~\cite{Gripaios:2016mmi} was proposed based on a classification given by Weinberg and d'Hoker \cite{DHoker:1994rdl}. As we show in \S\ref{previous classifications}, this classification is invalid if $G$ is disconnected or if $\pi_d (G/H) \neq 0$.

A fuller discussion of these terms, as well as others arising in composite Higgs models, will be given in \cite{Davighi:2018xwn}.

\section{Computing the spaces of AB and WZ terms \label{compute}}

Now we would like to summarize our classification, and also to show how the computation of the space of possible terms may be achieved in a given case. The classification states that
there are two types of topological term in $p$-dimensional sigma models on $G/H$, subject to our physical assumptions of \S \ref{assumptions}, which we have classified for general $G/H$ (at least for connected $G$). These are:

\begin{enumerate}
\item Aharonov-Bohm (AB) terms, classified by $H^p(G/H,U(1))$, the $p$th singular cohomology of $G/H$ valued in $U(1)$. In this paper, we have only discussed AB terms corresponding to the free part of $H^p(G/H,U(1))$, which are classified by
the quotient of the $p$th de Rham cohomology  by  its integral subgroup:
\begin{equation}
H^p_{dR}(G/H,\mathbb{R})/H^p_{dR}(G/H,\mathbb{Z}). \label{ab group}
\end{equation}
\item Wess-Zumino (WZ) terms, classified by the space of closed, integral, $(p+1)$-forms on $G/H$ satisfying the Manton condition, that is
\begin{equation}
\{ \omega \in Z^{p+1}(G/H,\mathbb{Z}) \ | \quad \forall X\in \mathfrak{g}\ \exists f_X\in \Lambda^{p-1}(G/H) \ \text{s.\ t.\ }  i_X(\omega)=df_X\}, \label{wz group}
\end{equation}
where $Z^{p+1}(G/H,\mathbb{Z})$ is the space of closed, integral $(p+1)$-forms.
\end{enumerate}
As we have seen, both the spaces of AB terms and WZ terms have the structure of an abelian Lie group; addition in the group corresponds to addition of the associated actions (or, equivalently, multiplication of the $U(1)$-valued action phases). 

We now turn to the question of how to compute these two groups in a given case. The group of (torsionless) AB terms (\ref{ab group}) is relatively easy to compute, being directly related to de Rham cohomology, for which a variety of tools are available. One of those, which is especially pertinent here, is that when $M\cong G/H$ and $G$ is connected and compact, the $p$th de Rham cohomology is isomorphic to the Chevalley-Eilenberg cohomology \cite{0031.24803} obtained from the complex of $G$-invariant $p$-forms on $G/H$ under the exterior derivative $d$. 

This complex is, moreover, useful for the computation of WZ terms, because they arise as a subspace of the 
closed $G$-invariant forms on $G/H$ in degree $p+1$.

So, how do we compute the $G$-invariant $q$-forms on $G/H$ (where we are interested in $q$ being $p$ or $p+1$)? Starting from the Maurer-Cartan form on $G$ itself, we form left-invariant $q$-forms on $G$ by choosing a basis for the Lie algebra and taking $q$-fold wedge products of the basis 1-forms. From these forms, one constructs well-defined (and $G$-invariant) $q$-forms on $G/H$ by restricting to the subset $\{ \Omega\}$ which are projectable onto $G/H$, that is, those $\Omega$ for which there is a unique $q$-form $\bar{\Omega}$ on $G/H$ which pulls back to $\Omega$ under the canonical projection onto cosets.

At least if $H$ is connected, the algorithm simplifies further to a computation at the level of the Lie algebras of $G$ and $H$, in that projectability is guaranteed by the local conditions $L_{Y}\Omega=0$ and $\iota_{Y}\Omega=0$, for all vector fields $Y$ on $G$ generating \textit{right} $H$ transformations ~\cite{0031.24803}.
In this case, the cohomology of such forms under $d$ is isomorphic to the relative Lie algebra cohomology of $\mathfrak{g}$ with respect to $\mathfrak{h}$ ~\cite{0031.24803}. 

Thus, if $G/H$ is compact and $H$ is connected, we may compute the space of AB terms algebraically, by finding the $p$th relative Lie algebra cohomology, and quotienting by integral classes. Moreover, given only that $H$ is connected ($G/H$ may now be non-compact), we may compute the space of WZ terms by finding the space of $(p+1)$-cocycles in the relative Lie algebra cohomology (over integers),\footnote{An important distinction to note is that, unlike the AB group, the possible WZ terms are properly classified by cocycles, not cohomology classes. Nevertheless, because these cocycles are a subspace of the space of $G$-invariant forms on $G/H$, they are guaranteed to form a subspace of a finite-dimensional vector space. Thus, even in the worst-case scenario, the computation of the space of topological terms can be carried out in an algorithmic fashion.} and then restricting to the subset that satisfy the Manton condition. This last step is not, in general, reducible to algebra.

How then, in practice, does one enforce the Manton condition? In fact, the Manton condition is automatically satisfied for all vector fields $X\in [\mathfrak{g},\mathfrak{g}]\subset \mathfrak{g}$, and so need only be checked for generators of the Abelianization of $\mathfrak{g}$, that is the quotient $\mathfrak{g}/[\mathfrak{g},\mathfrak{g}]$. The proof is as follows. For each vector field  $X\in [\mathfrak{g},\mathfrak{g}]$, one can write $X=[Y,Z]$, for $Y,\ Z$ also in $\mathfrak{g}$. This, together with the identity $[L_Y, \iota_Z] \alpha = \iota_{[Y,Z]} \alpha$ (where  $\alpha$ is any differential form), implies that 
\begin{equation}
\iota_X \omega = \iota_{[Y,Z]}\omega = L_Y\iota_Z\omega=d (\iota_Y \iota_Z \omega), \label{semi simple}
\end{equation}
where in the second equality we used $L_Y\omega=0$, and in the final equality we used $L_Y=\iota_Y d+d\iota_Y$ and that $d(\iota_Z \omega)=0$. This proves the claim. Furthermore, this argument gives us an explicit construction for the Noether current $(p-1)$-forms associated with those vector fields $X\in [\mathfrak{g},\mathfrak{g}]$; we simply contract 
$\omega$ with two vector fields $Y$ and $Z$ whose Lie bracket is $X$. We find it striking that a local version of the result (\ref{semi simple}) was formulated by Manton and collaborators, in the context of spacetime symmetries of gauge theories \cite{Forg_cs_1980,JACKIW1980257}; however, considerations of the global topology of $G$ and $G/H$, which have been central to the present work (most evidently in the formulation of the Manton condition), were not considered there.

As an important corollary, if $G$ is a semi-simple Lie group ({\em i.e.} when $\mathfrak{g} = [\mathfrak{g},\mathfrak{g}]$), the Manton condition for $G$-invariance is necessarily satisfied for any $G$-invariant form; thus, in this case, the computation of the space of WZ terms indeed reduces to algebra (assuming only connectedness of $H$).

Finally, we address the subtlety that arises when the subgroup $H$ is disconnected.
When $H$ is disconnected, one can no longer restrict to the subset $\{ \Omega\}$ of $G$-invariant forms that are projectable to $G/H$ using only local conditions (at the level of the Lie algebra). Rather, one must check in addition that the putatively projectable form on $G$ is in fact invariant under the group of disconnected components of $H$.
As an example in $p=1$ of the consequences of disconnected $H$ for our classification, consider the difference between quantum mechanics on $S^2\cong SO(3)/SO(2)$ {\em vs.} $\mathbb{R}P^2\cong SO(3)/O(2)$, the real projective plane. 
The first case corresponds to the Dirac monopole, and there is a WZ term as we have discussed, which can be established using the conditions above at the level of the Lie algebra alone. But despite the fact that $O(2)$ and $SO(2)$ have the same Lie algebra, there is no WZ term for $SO(3)/O(2)$,
for the simple reason that  any candidate $SO(3)$-invariant 2-form 
 must be proportional to the volume form, 
and there is no volume form on the non-orientable manifold $\mathbb{R}P^2$.

The reader may have noticed that of the examples we have discussed so far, none have featured both AB and WZ terms. It is nonetheless easy to construct examples which do. For example, consider quantum mechanics on $G/H=\mathbb{R}^3/\mathbb{Z}\cong S^1\times\mathbb{R}^2$, for which the AB group is $\mathbb{R}/\mathbb{Z}\cong U(1)$ and the WZ group is $\mathbb{R}$. A highly non-trivial example featuring both AB and WZ terms is provided by a Composite Higgs theory based on the coset $G/H=SO(6)/SO(4)$. We shall describe the topological terms in this model in \cite{Davighi:2018xwn}.

\subsection{Comparison with previous classifications \label{previous classifications}}

We have already given some indication of how our homological approach to topological terms differs from a homotopic approach (which applies only for worldvolumes that are homeomorphic to $p$-dimensional spheres).
In this Section, we
comment in more detail on how
our classification compares with previous partial classifications of topological terms presented in ~\cite{DHoker:1994rdl} and  ~\cite{dijkgraaf1990}.

The classification by Weinberg and d'Hoker in \cite{DHoker:1994rdl} is based on homotopy and purports to apply to arbitrary $G/H$, provided only that $G$ is compact. The claim is that WZ terms (defined there as terms in the lagrangian which shift by a non-vanishing total derivative under the $G$-action) are in one-to-one correspondence with the (p+1)th de Rham cohomology of $G/H$.

It is claimed in \cite{DHoker:1994rdl} that when the sigma model map $\phi:S^p\rightarrow G/H$ is not homotopic to a constant map, one can nevertheless define the action as the sum of two pieces, as follows. One piece is an action assigned to any one fixed representative in each homotopy class; the other piece is the integral (as in the Witten construction) of a closed $(p+1)$-form over a $(p+1)$-dimensional submanifold (call it $N$) defined by a homotopy linking the map $\phi$ to the fixed representative.

This prescription is not only somewhat cumbersome (especially in cases where there are infinitely many homotopy classes), but also leads to problems with $G$-invariance, as we now discuss.

Let us start by considering the closed $(p+1)$-form. It is claimed in \cite{DHoker:1994rdl} that `The group $G$ acts transitively on the manifold $G/H$, so a $G$ transform of a form define[s] the same de Rham cohomology class.' The simplest example that shows this claim to be false in general is given by $G = \mathbb{Z}/2\mathbb{Z} $ acting on itself. We have that $H^0_{dR} (\mathbb{Z}/2\mathbb{Z}) = \mathbb{R}^2$, whose 2 generators may be represented by the $0$-forms taking value unity on one component and vanishing on the other. The $G$-action does not send these forms (nor their classes) into themselves, but rather interchanges them.\footnote{To give a more physically-relevant example, the classification given in \cite{DHoker:1994rdl} also yields the wrong answer for a non-minimal composite Higgs model based on $G/H = O(6)/O(5)\cong S^5$,  featuring custodial protection of $Z \rightarrow b\overline{b}$. 
Elements in $O(6)$ that are disconnected from the identity send the volume form (and hence the de Rham class) to minus itself, such that there is no $O(6)$-invariant topological term. We discuss invariance of AB and WZ terms under disconnected groups in Appendix \ref{disconnected}. } What is true is that the action of any $g \in G$ on $G/H$ (or indeed on any manifold on which it acts) is a diffeomorphism of $G/H$ which induces an automorphism on de Rham cohomology and that when $g$ is connected to the identity the diffeomorphism is homotopic to the identity map and so induces the identity automorphism on de Rham chomology, sending each class into itself.

Thus the specific claim in \cite{DHoker:1994rdl} would be valid if one additionally assumes that $G$ is {\em connected}. But even this further restriction is not enough to guarantee $G$-invariance of the action, because the action of $G$ on $G/H$ moves the image of the worldvolume, but not the fixed representative. Therefore, the $G$-action results in a new submanifold $N'$, which is not the one induced from $N$ by the action of $G$ on $G/H$. As a result, $G$-invariance of the $(p+1)$-form does not guarantee invariance of the action.

This problem invalidates the classification given in \cite{DHoker:1994rdl}  when $\pi_{p} (G/H) \neq 0$, and it is far from clear how to fix it in a homotopy-based approach. But from the homological perspective, the problem is already fixed: a topological term is possible iff. the Manton condition (which is stronger than the condition of $G$-invariance of $\omega$) is satisfied. Moreover, this condition is also valid for non-compact $G$.

Our examples of quantum mechanics on the torus and the composite Higgs model based on $SO(5)\times U(1)/SO(4)$, where the relevant homotopy groups are non-vanishing, show that, in many cases, the classification in \cite{DHoker:1994rdl} suggests the existence of a WZ term when in fact there is none. But it is also quite possible that there do exist WZ terms even when $\pi_{p} (G/H) \neq 0$. Good candidates for $G/H$ are those for which $\pi_{p} (G/H) \neq 0$ but $G$ is semi-simple, such that the Manton condition is implied by $G$-invariance of $\omega$. The Composite Higgs theory with $p=4$ and $G/H=SO(6)/SO(4)$ (for which $\pi_4=\mathbb{Z}$), provides such an example.

Turning to the other partial classification, it is claimed in a paper by Dijkgraaf and Witten~\cite{dijkgraaf1990} that topological terms in a $p=2$ sigma model with target space being a compact group $G$ (not necessarily connected or simply connected), that are invariant under the left-right action by $G\times G$, are classified by $H^3(G,\mathbb{Z})$. Such theories with two-sided $G$-invariance are appropriately termed `chiral theories'.

One can see that our classification agrees with that of Dijkgraaf and Witten \cite{dijkgraaf1990} in the case where $G$ is semi-simple. In this case $G\times G$ is also semi-simple, and thus the Manton condition is necessarily satisfied.\footnote{In fact, when the target space is a Lie group $G$, as it is here, $G$ being semi-simple implies $H^2(G,\mathbb{R})=0$, such that {\em any} closed 2-form is necessarily exact.} The space of WZ terms is thence given by the space of closed, integral, bi-invariant $3$-forms. Because $G$ is a symmetric space, every bi-invariant form is closed \cite{schwarz1994}, hence the closed, integral, bi-invariant $3$-forms are in one-to-one correspondence with the integral cohomology classes in the Chevalley-Eilenberg cohomology of $G\times G$ relative to $G$. Since $G$ is assumed compact, there is an isomorphism between the Chevalley-Eilenberg cohomology and the de Rham cohomology of $(G \times G)/G \cong G$ \cite{0031.24803}. Thus, the WZ terms are in one-to-one correspondence with the integral de Rham cohomology classes of $G$ in degree $3$.
Our classification also contains, in general, AB terms, but these vanish because $H^2(G,\mathbb{R})=0$. There is, however, a contribution coming from the torsion subgroup of $H^2(G,\mathbb{Z})$, 
which we have neglected in our classification because we insisted on globally-defined AB terms; torsion terms correspond to {\em locally}-defined AB terms (see Appendix \ref{local aharonov bohm}).\footnote{
An equivalent but perhaps more elegant way to incorporate torsion is through a classification based on differential characters \cite{JDBGORW}.}  When torsion is included (through the locally-defined AB terms), we find that the full space of topological terms is given by $H^3 (G, \mathbb{Z})$, in agreement with Dijkgraaf and Witten.

When $G$ is not semi-simple, Dijkgraaf and Witten claim that a topological term is given by ``any differential character'' and that the space of such terms contains extra pieces ``corresponding to generalized $\theta$ angles on the torus $H^2(G,\mathbb{R})/\rho(H^2(G,\mathbb{Z}))$''. We certainly agree with the second claim, since the generalized $\theta$ angles are just our AB terms. But we do not agree with the first part of the claim, because it neglects the requirement of $G$-invariance. The differential characters include those corresponding to all closed, integer, 3-forms, whereas in fact only those satisfying the Manton condition lead to a $G$-invariant action. Given our discussion in \S \ref{compute}, we see that it remains to check the Manton condition on $\mathfrak{g}/[\mathfrak{g},\mathfrak{g}]$.

A simple example should suffice to highlight the discrepancy.    
Let $G=U(1)^3$, for which $H^3(G,\mathbb{Z})=H^3(T^3,\mathbb{Z})=\mathbb{Z}$, generated by the 3-form $\omega=dx\wedge dy\wedge dz$ which integrates to unity over the 3-torus (where $x\sim x+1$, $y\sim y+1$, and $z\sim z+1$). Exactly as we have seen for quantum mechanics on the 2-torus, the Manton condition fails for each vector field generating $G$, and one cannot write down a $G$-invariant WZ term. Explicitly, the problem is that one cannot write down an invariant action for cycles corresponding to non-trivial classes in $H_2(T^3,\mathbb{Z})$, corresponding to toroidal worldsheets.

\section{Discussion \label{conclusion}} 

We have classified, from a homological perspective, the space of topological terms for a generic non-linear sigma model on a homogeneous space $G/H$, assuming those terms can be written in terms of local differential forms. We divide the space of such terms into two parts, corresponding to contributions from (local) $p$-forms which are either closed or not closed, and we have called these AB and WZ terms respectively. At least for connected $G$, and neglecting torsion, the physically-inequivalent AB terms are in one-to-one correspondence with the $p$th de Rham cohomology of $G/H$, quotiented by the subgroup of integral classes. The WZ terms are in one-to-one correspondence (for connected $G$) with those closed, integral, $(p+1)$-forms whose interior products with the generators of the $G$-action are exact $p$-forms. This condition for $G$-invariance, which we call the Manton condition, was derived by requiring $G$-invariance of the action on all (smooth singular) homology $p$-cycles, which implies that we allow worldvolumes of arbitrary topology (subject to the requirement that they be smooth, orientable manifolds of dimension $p$).

Our formulation of topological terms started from representing worldvolumes by homology cycles. In general, homology groups contain both free and torsion subgroups, and in this paper we have neglected the torsion part by insisting on AB terms corresponding to only globally-defined, closed $p$-forms.
It is natural to ask whether such torsion elements in homology can also give rise to topological terms in sigma models on $G/H$.

The answer is that they can, as we illustrate by means of an example. 
The quantum mechanics of a rigid body may be described by a worldline on $SO(3) \cong \mathbb{R}P^3$ (representing the configuration space of the rigid body as described by, say, its Euler angles). The first homology group of $\mathbb{R}P^3$ is isomorphic to $\mathbb{Z}/2 \mathbb{Z}$. At the level of the fundamental group (which, by Hurewicz' theorem, is isomorphic to the first homology group), the non-trivial element may be represented by a loop connecting a point in $S^3 \cong SU(2)$ to its antipode (which becomes a closed loop in $SO(3)$ once we identify). Evidently, we can define two distinct topological action phases by either associating a phase of unity to all loops (or cycles) or associating unity to trivial cycles and negative unity to non-trivial cycles. The physical interpretation is that, in the former case, the rigid body is bosonic, while in the latter case it is fermionic.
While such torsion effects can indeed be described with locally-defined differential forms, there is an elegant formalism, using differential characters, which captures these torsion effects (in addition to the AB and WZ terms which we have considered in this paper). The details will be described in a forthcoming work \cite{JDBGORW}.

Even with such terms added, it is known that there exist yet more topological terms for sigma models on homogeneous spaces, which cannot be captured by a homological classification. To give just one example, consider a worldvolume homeomorphic to $S^4$ and $G/H=SU(2)\cong S^3$. Since $\pi_4(S^3)=\mathbb{Z}/2\mathbb{Z}$, there are two homotopy classes of maps and one may define a non-trivial $\mathcal{O}$-invariant action phase by assigning a phase of $-1$ to maps $\phi :S^4 \rightarrow SU(2)$ in the non-trivial homotopy class. Since $H_4(S^3)= 0$, it is clear that such terms cannot be captured by a homological classification. The physics of such a term is as follows \cite{Finkelstein:1968hy,Witten:1983tx}. Since $\pi_3(S^3)=\mathbb{Z}$, the theory contains solitons. A map $\phi$ corresponding to a process in which a solition-antisoliton pair is created and the soliton is rotated by $2\pi$ before the pair annihilates lies in the non-trivial homotopy class. The topological term may thus be interpreted as assigning fermionic character to the solitons of the theory.

\acknowledgments

We are very grateful to Christian B\"{a}r, Nakarin Lohitsiri, Oscar Randal-Williams, David Tong, and Bryan Webber for discussions. BG is partially supported by STFC consolidated grant ST/P000681/1 and King's College, Cambridge. JD is supported by The Cambridge Trust and STFC consolidated grant ST/P000681/1.

\appendix
\section{Locally-defined Aharonov-Bohm terms and torsion} \label{local aharonov bohm}

Suppose an Aharonov-Bohm term is defined from a set of closed $p$-forms which are at first defined only locally on patches. How do we classify the corresponding physically-inequivalent theories?

We content ourselves with an explicit analysis of the case $p=1$; generalization is straightforward, but tedious. Thus, we take $\{A^1_{\alpha}\}$ to be a collection of closed 1-forms defined on the open sets $\{U_{\alpha}\}$, satisfying $A^1_{\alpha}-A^1_{\beta}=dA^0_{\alpha\beta}$ on double intersections. We define the action to be the $p=1$ specialization of (\ref{WYA}), regarding this locally-defined AB term as a WZ term for which the $(p+1)$-form $\omega=0$.
Closure of $\{A^1_{\alpha}\}$ means we can write $A^1_{\alpha}=dB_{\alpha}$ using the Poincar\'e lemma, for local 0-forms $B_{\alpha}$. Define a set of 0-forms $\chi_{\alpha\beta}$ for each double intersection by
\begin{equation}
\chi_{\alpha\beta}=B_{\alpha}-B_{\beta}-A^0_{\alpha\beta} \quad \textrm{on}\  U_{\alpha\beta}, \label{chi}
\end{equation}
which can be written in terms of the \v Cech coboundary operator as $\{\chi\}=\delta\{B\}-\{A^0\}$, and are evidently antisymmetric.
Using $\{A^1\}=\{dB\}$ and $\delta\{A^1\}=d\{A^0\}$, we have that
$d\chi_{\alpha\beta}=0$ throughout $U_{\alpha\beta}$,
that is, the 0-forms $\chi_{\alpha\beta}$ are in fact constants on double intersections.
Evaluating the action on a cycle $z$ whose image intersects three double intersections $U_{\alpha\beta}$, $U_{\beta\gamma}$, and $U_{\gamma\alpha}$, gives
\begin{equation}
S[z]=\chi_{\alpha\beta}(c_{0,\alpha\beta}) + \chi_{\beta\gamma}(c_{0,\beta\gamma}) + \chi_{\gamma\alpha}(c_{0,\gamma\alpha}), \label{local ab action}
\end{equation}
where $c_{0,\alpha\beta}$ \&{\em c.} are 0-chains whose images are in the double intersections. Thus, the action is
a sum over constants defined on double intersections.

The constant-valued \v Cech 1-cochain $\{\chi\}$ is furthermore closed under the \v Cech coboundary operator $\delta$. To see why, suppose our good cover on $M$ contains non-vanishing triple intersections, and consider a cycle $z^*$ whose image is contained in a particular $U_{\alpha\beta\gamma}$. We can therefore choose to write $z^*=c_{1,\alpha}$ and integrate $A^1_{\alpha}$ over the whole cycle, which yields $S[z^*] = 0$ because $A^1_{\alpha}$ is closed and therefore exact on $U_{\alpha}$.
But, the action is invariant under repeated subdivision of our cycle, and so we could subdivide $z^*$ until it consists of at least three 1-chains, and then choose to write $\mathrm{Sd}^n z^*=c_{1,\alpha}+c_{1,\beta}+c_{1,\gamma}$, with the boundaries between these 1-chains being a set of three 0-chains. 
The action then evaluates to $S[z^*]=\chi_{\alpha\beta}+\chi_{\beta\gamma}+\chi_{\gamma\alpha}$ which must therefore equal an integer, since the two computations must agree on the value of $e^{2\pi i S[z]}$. This implies
\begin{equation}
\delta\{\chi_{\alpha\beta}\}\in \mathbb{Z},
\end{equation}
and hence $\{\chi_{\alpha \beta}\}$ is a \v Cech 1-cocycle when it is regarded as having values in $\mathbb{R}/\mathbb{Z}\cong U(1)$ and therefore defines a cohomology class $[\{\chi_{\alpha \beta}\}] \in H_{\check{c}}^1(M,U(1))$.\footnote{Of course, if our manifold doesn't have any triple intersections, any \v Cech 1-cochain is trivially a 1-cocycle because there are no nonzero \v Cech 2-cochains.}
Moreover, the action only depends on this cohomology class, because it vanishes identically when evaluated on any \v Cech 1-coboundary ({\em i.e.} if $\{\chi_{\alpha\beta}\}=\delta\{e_{\alpha}\}=\{e_\alpha - e_{\beta}\}$, where $e_{\alpha}$ are constant on the open sets $U_\alpha$).
The space of such locally-defined AB actions is therefore isomorphic to the 1st \v Cech cohomology with coefficients in $U(1)$.

Moreover, since $M$ is a smooth manifold, \v Cech cohomology is isomorphic to singular homology, for which we have the short exact sequence $$0 \rightarrow \Omega^1_{cl} (M) /\Omega^1_0 (M) \rightarrow H^1 (M,U(1))\rightarrow \mathrm{Ext} (H_1 (M,\mathbb{Z}),\mathbb{Z}) \rightarrow 0,$$ where $\Omega^1_{cl} (M)$ denotes the closed 1-forms on $M$, $\Omega^1_0 (M)$ denotes the closed, integral 1-forms on $M$, and $\mathrm{Ext}$ accounts for torsion. Thus, ignoring the torsion, the AB terms can be represented by globally-defined, closed 1-forms.

Explicitly, the action can be written as follows. 
Because $\{\chi\}$ is both \v Cech and de Rham closed, following Alvarez \cite{Alvarez:1984es} we can ``invert" $\delta$ to construct a set of 0-forms $\tilde{B}_{\alpha}$ out of the constants $\chi_{\alpha\beta}$ such that $\tilde{B}_{\alpha}-\tilde{B}_{\beta}=\chi_{\alpha\beta}$ on each double overlap (explicitly, $\tilde{B}_{\alpha} = \sum_{\gamma} \chi_{\alpha\gamma} p_{\gamma}$, where the collection of functions $\{p_{\gamma}\}$ is a partition of unity). That $\delta\{\tilde{B}\}=\{\chi\}$ follows only from the antisymmetry of $\chi_{\alpha\beta}$ and the cocycle condition.  Then define a new set of local 1-forms $\{\tilde{A}^1_\alpha\}=\{d\tilde{B}_\alpha\}$. These satisfy $\delta\{\tilde{A}^1\}=\delta\{d\tilde{B}\}=d\delta\{\tilde{B}\}=d\{c\}=0$ using the commutativity of $d$ and $\delta$. Hence the local 1-forms $\{\tilde{A}^1\}$ in fact agree on double overlaps and therefore define a global, closed 1-form. The WZ action on the cycle $z$ for this globally-defined 1-form is
\begin{equation}
\int_z \tilde{A}^1 = \int_{c_{1,\alpha}} \tilde{A}^1_{\alpha} + \int_{c_{1,\beta}} \tilde{A}^1_{\beta} + \int_{c_{1,\gamma}} \tilde{A}^1_{\gamma} = (\tilde{B}_{\alpha} - \tilde{B}_{\beta})(c_{0,\alpha\beta}) + (\tilde{B}_{\beta} - \tilde{B}_{\gamma})(c_{0,\beta\gamma}) + (\tilde{B}_{\gamma} - \tilde{B}_{\alpha})(c_{0,\gamma\alpha}),  
\end{equation}
which sure enough equals (\ref{local ab action}). The upshot is that, given locally-defined closed 1-forms $\{A^1\}$, the WZ action just evaluates to
\begin{equation}
S_{WY}[z]=\int_z \tilde{A}^1,
\end{equation}
for the global 1-form $\tilde{A}^1$ constructed out of $\{A^1\}$ via a partition of unity as above, which is the action for an AB term (\ref{ab action}). This argument generalizes to any cycle, and higher $p$.

\section{Consistency of the Wess-Zumino action phase} \label{ambiguities}

As we described in \S \ref{consistency}, the action for a WZ term is written as a sum of integrals of locally-defined forms (which are constructed from a closed $(p+1)$-form $\omega$) over $\mathcal{U}$-small chains  of the appropriate degree, and contained within the appropriate intersections of open sets (which are constructed from the worldvolume cycle by repeated subdivision). In this Appendix, we show that the action (\ref{WYA}) constructed in this way is free of any ambiguities that might arise when there is a choice of locally-defined forms to integrate on a given chain.

First consider a $p$-simplex $\sigma$ which is contained in a double intersection $U_{\alpha\beta}$, and on which we can therefore integrate either $A^p_{\alpha}$ or $A^p_{\beta}$. 
The boundary of $\sigma$ is the sum of two $(p-1)$-chains, which we denote $e_{\alpha}$ and $e_{\beta}$ (that is $\partial\sigma=e_{\alpha}+e_{\beta}$), which originate from taking the boundary of $c_{p,\alpha}$ and $c_{p,\beta}$ respectively.
If we choose to integrate $A^p_{\alpha}$ on $\sigma$, the relevant pieces of the action are
\begin{equation}
S_{\alpha}=\int_{\sigma} A^p_{\alpha} - \int_{e_{\beta}} A^{p-1}_{\alpha\beta}. \label{S alpha}
\end{equation}
If we choose to integrate $A^p_{\beta}$ on $\sigma$, the relevant pieces of the action are
\begin{equation}
S_{\beta}=\int_{\sigma} A^p_{\beta} - \int_{e_{\alpha}} A^{p-1}_{\beta\alpha}. \label{S beta}
\end{equation}
The difference is 
\begin{equation}
S_{\alpha}-S_{\beta}=\int_{\sigma} (A^p_{\alpha}-A^p_{\beta}) - \int_{\partial \sigma}A^{p-1}_{\alpha\beta}=\int_{\sigma} (A^p_{\alpha}-A^p_{\beta} - dA^{p-1}_{\alpha\beta}), \label{ambiguity 1}
\end{equation}
where in the second equality we have used Stokes' theorem.  Hence, the ambiguity vanishes if $\{dA^{p-1}_{\alpha\beta}\}=\delta\{A^p_{\alpha}\}$, as encoded in the tic-tac-toe table (\ref{QFT TTT}).

\begin{figure}
\begin{center}
\begin{tikzpicture}
  \draw[black] (-5,-3) -- (-1,-1) -- (0,1) -- (0,4);
  \fill[red!12!white] (-5,-3) -- (-1,-1) -- (0,1) -- (0,4) -- (-5,4);
  \draw[black] (5,-3) -- (1,-1) -- (0,1) -- (0,4);
  \fill[blue!12!white] (5,-3) -- (1,-1) -- (0,1) -- (0,4) -- (5,4);
  \draw[black] (-5,-3) -- (-1,-1) -- (1,-1) -- (5,-3);
  \fill[orange!12!white] (-5,-3) -- (-1,-1) -- (1,-1) -- (5,-3);
  
  \node[label] at (-4,3)  {\Large $c_{\alpha}$};
  \node[label] at (4,3)  {\Large $c_{\beta}$};
  \node[label] at (-2,-2.5)  {\Large $c_{\gamma}$};
  \node[label] at (0,-0.25)  {\Large $\sigma$};

  \node[label] at (-0.75,0.1)  {$e_{\alpha}$};
  \node[label] at (0.75,0.1)  {$e_{\beta}$};
  \node[label] at (0.0,-1.2)  {$e_{\gamma}$};

  \node[label] at (0.2,1.1)  {$C$};
  \node[label] at (1.2,-0.8)  {$A$};
  \node[label] at (-1.2,-0.8)  {$B$};
  
  \draw[black,  thick] (-1,-1) -- (0,1) -- (1,-1) -- cycle;

  \draw[red] (3,4) .. controls (2.5,-1.5) .. (-1,-3);
  \draw[blue] (-3,4) .. controls (-2.5,-1.5) .. (1,-3);
  \draw[orange] (-5,-1) .. controls (0,2.5) .. (5,-1);

\end{tikzpicture}
\caption{In $p=2$, there is a potential ambiguity in the action when a 2-simplex $\sigma$ in our $\mathcal{U}$-small chain complex lies in a triple intersection of open sets. In this diagram, $U_{\alpha}$ is the region to the left of the curved red line, such that $\mathrm{Im}\ c_{\alpha} \subset U_{\alpha}$, and $U_{\beta}$ ($U_{\gamma}$) are the regions to the right of (below) the curved blue (orange) lines respectively. The 0-, 1-, and 2-chains depicted are labelled as in the main text.
}
\label{triple intersection}
\end{center}
\end{figure}
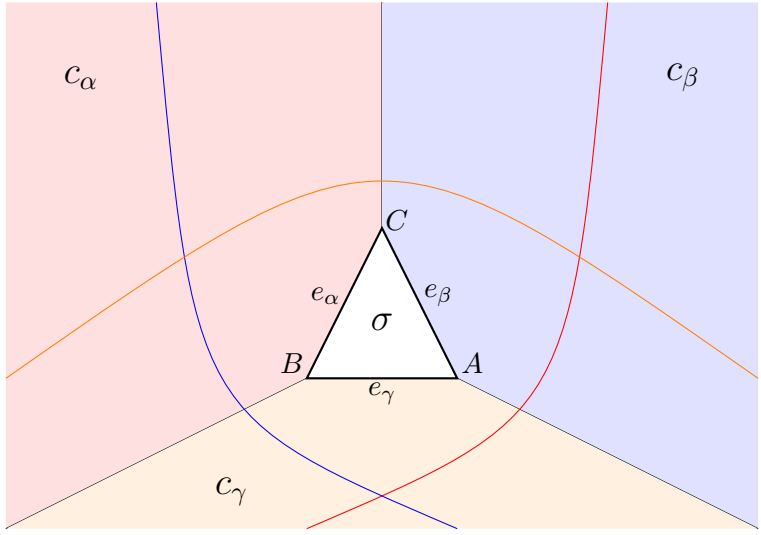

However, as we anticipated above, there are further ambiguities. Suppose there exists a $p$-simplex $\sigma$ which is contained not just in a double intersection, but in a triple intersection of open sets, $U_{\alpha\beta\gamma}$, and on which we can therefore integrate $A^p_{\alpha}$, $A^p_{\beta}$, or $A^p_{\gamma}$.
We suppose that $c_{\alpha}$, $c_{\beta}$, and $c_{\gamma}$ all intersect 
$U_{\alpha\beta\gamma}$, and that the boundary $\partial \sigma$ is thus now the sum of three $(p-1)$ chains, {\em viz.} $\partial\sigma=e_{\alpha}+e_{\beta}+e_{\gamma}$, each originating from the boundary of $c_{p,\alpha}$, $c_{p,\beta}$, and $c_{p,\gamma}$.  To be concrete, let us consider the case $p=2$, in which case $\sigma$ is a 2-simplex at which the $\mathcal{U}$-small 2-chains $c_{\alpha}$, $c_{\beta}$, and $c_{\gamma}$ meet, and $e_{\alpha}$, $e_{\beta}$, and $e_{\gamma}$ are 1-chains whose sum is $\partial \sigma$. The boundaries of these 1-chains are themselves three 0-chains ({\em i.e.} points), call them $A$, $B$, and $C$, corresponding to the vertices of the 2-simplex $\sigma$. Specifically, let $A$ be the point common to $\partial e_{\beta}$ and $\partial e_{\gamma}$, let $B$ be the point common to $\partial e_{\gamma}$ and $\partial e_{\alpha}$, and $C$ be the point common to $\partial e_{\alpha}$ and $\partial e_{\beta}$. The situation is depicted in Fig. \ref{triple intersection}.

If we choose to integrate, respectively, $A^2_{\alpha},A^2_{\beta}$ or $A^2_{\gamma}$ on $\sigma$, the relevant pieces of the action are, respectively,
\begin{gather}
\begin{aligned}
& S_{\alpha}=\int_{\sigma} A^2_{\alpha} - \int_{e_{\beta}} A^{1}_{\alpha\beta} - \int_{e_{\gamma}} A^{1}_{\alpha\gamma}+ A^{0}_{\alpha\beta\gamma}(A), \\
& S_{\beta}=\int_{\sigma} A^2_{\beta} - \int_{e_{\gamma}} A^{1}_{\beta\gamma} - \int_{e_{\alpha}} A^{1}_{\beta\alpha}+ A^{0}_{\beta\gamma\alpha}(B), \\
& S_{\gamma}=\int_{\sigma} A^2_{\gamma} - \int_{e_{\alpha}} A^{1}_{\gamma\alpha} - \int_{e_{\beta}} A^{1}_{\gamma\beta}+ A^{0}_{\gamma\alpha\beta}(C).
\end{aligned} \label{ambiguity 2}
\end{gather}
The difference between, say, $S_{\alpha}$ and $S_{\beta}$ is
\begin{equation}
S_{\alpha}-S_{\beta}=\int_{\sigma} (A_{\alpha}-A_{\beta}) - \int_{e_{\gamma}} (A_{\alpha\gamma}+A_{\gamma\beta}) - \int_{\partial\sigma - e_{\gamma}} A_{\alpha\beta} + A_{\alpha\beta\gamma}(A) - A_{\alpha\beta\gamma}(B)
\end{equation}
(where we have suppressed the superscripts indicating the degree of the forms). This is equal to 
\begin{equation}
S_{\alpha}-S_{\beta}=\int_{\sigma} (A_{\alpha}-A_{\beta} - dA_{\alpha\beta})
-\int_{e_{\gamma}} (A_{\alpha\gamma}+A_{\gamma\beta}+A_{\beta\alpha}-dA_{\alpha\beta\gamma}),
\end{equation}
where Stokes' theorem has been used twice, noting that $A-B=\partial e_{\gamma}$ (we obtain a permutation of this expression for each pairwise difference of the three actions in (\ref{ambiguity 2})). The first term is guaranteed to vanish given we have removed the ambiguity in (\ref{ambiguity 1}). Hence, this second ambiguity due to triple intersections vanishes, in general $p$,  when $\{dA^{p-2}_{\alpha\beta\gamma}\}=\delta\{A^{p-1}_{\alpha\beta}\}$, again as encoded in the tic-tac-toe table (\ref{QFT TTT}).

In a similar way, the tower of terms that we have included in the action, and the tic-tac-toe relations between them (\ref{QFT TTT}), are such that there are no ambiguities over which form to integrate at {\em any} degree greater than zero, with the ambiguity in forms of a given degree being removed by the presence of forms of one degree lower. In the case of general $p$, schematically, one has to remove ambiguities arising from $p+1$ diagrams, where in the $q$th diagram we consider the ambiguities in our definition of the action when a $p$-simplex is contained in a $(q+1)$-fold intersection, for $q=1,...,p+1$. For this $q$th diagram, there will be $q+1$ possible ways of writing the action, and insisting that their differences vanish thus yields $q$ independent constraints; $(q-1)$ of these constraints will be satisfied by the conditions that arise from the preceding $(q-1)$ diagrams (which will all be successive relations from the tic-tac-toe table), with the final $q$th constraint being that $\{dA^{p-q}\}=\delta\{A^{p-q+1}\}$.

\section{The case of disconnected $G$} \label{disconnected}

In this Appendix we discuss, as a somewhat technical aside, how the conditions for $G$-invariance of both AB and WZ terms must be modified when $G$ is a disconnected Lie group. We first discuss the story for AB terms, and then WZ terms.

\subsection{AB terms}

Let $G_0$ be the normal subgroup of $G$ given by the maximal component connected to the identity in $G$. The group of components $G/G_0$ is then a discrete group. A $G_0$-invariant AB term (constructed as in \S \ref{ab}) will be $G$-invariant iff.\ the corresponding closed $p$-form $A$ shifts by an exact form under the action of $G/G_0$, by (\ref{dR}). Indeed, the action on a cycle $z$, $S[z]=\int_z A$, shifts to $\int_z L_{gG_0}^* A$ under the action of $gG_0 \in G/G_0$, where in general $L_g^*$ denotes the action of $G$ that is induced on forms via pullback of the action of $g \in G$ on $G/H$. So the action phase will be invariant iff.\ $\int_z (L_{gG_0}^* -1)A \in \mathbb{Z}$ for all $z$ and for all $gG_0 \in G/G_0$. 

This condition is inequivalent to the (stronger) condition that $A$ be $G/G_0$-invariant, as the following example shows. 
Let $G/H = O(2)/O(1) \cong S^1$. The action of the non-trivial element in $G/G_0 = O(2)/SO(2) \cong \mathbb{Z}/2\mathbb{Z}$ sends the $SO(2)$-invariant 1-form $bdx$ to minus itself. Hence, we obtain an $O(2)$-invariant action upon integrating $bdx$ iff. $2b$ is integral (despite $bdx$ not being an $O(2)$-invariant 1-form). But if $b$ is integral, then we obtain a trivial action phase. Hence the space of AB terms for $O(2)/O(1)$ are isomorphic with the group $\mathbb{Z}/2\mathbb{Z}$, generated by $ dx/2$.

This example illustrates that, even though the group of components is finite, one cannot obtain the full set of $G$-invariant AB terms simply by averaging the $p$-form appearing in a $G_0$-invariant term with respect to the $G/G_0$-action.  Indeed, in this example, averaging any such $p$-form yields $0$.

\subsection{WZ terms}

Again let $G_0$ be the normal subgroup of $G$ given by the maximal component connected to the identity in $G$. For WZ terms, we have shown in \S \ref{consistency} that the integral over a boundary may be written in terms of the $(p+1)$-form $\omega$. So the action phase will be invariant only if $\int_b (L_{gG_0}^* -1) \omega \in \mathbb{Z}$ for all chains $b$. In fact, by the arguments of \S \ref{injectivity}, we have the stronger requirement that $\int_b (L_{gG_0}^* -1) \omega = 0$, for all chains $b$, which by de Rham's theorem (\ref{dR}) implies that $\omega$ must be $G$-invariant. Thus $G$-invariance of $\omega$ is a necessary condition.

Evidently, $G$-invariance of $\omega$ cannot be a sufficient condition, since it fails in the case where $G$ is in fact connected (in which case we need the stronger Manton condition if the action is to be invariant on all cycles, not just those which are boundaries). It also fails when $\omega = 0$, such that we are, in fact, describing an AB term. Indeed, we have already seen that $G$-invariance of AB terms is automatic only when $G$ is connected, and is otherwise non-trivial.

It is, however, possible to establish that, when $\omega$ is $G$-invariant, the shift in the corresponding topological term is itself a topological term, but of AB type. In other words, it is always possible to write the shift in the action on a $p$-cycle in terms of an integral of some closed, globally-defined $p$-form over the cycle. In particular, the shift due to $g \in G$ of an AB term described by $p$-form $A$ can be written as the integral of the closed $p$-form $(L_g^* -1)A$, and the infinitesimal shift of a WZ term described by $(p+1)$-form $\omega$ can be written as the integral of the closed $p$-form $\iota_X \omega$.

We postpone proof of this result, and exploration of its consequences, to \cite{JDBGORW}, contenting ourselves here with an illustrative example: consider
quantum mechanics ($p=1$) on $G/H=O(3)/O(2) \cong S^2$. The action of the non-trivial element in $G/G_0 = O(3)/SO(3) \cong \mathbb{Z}/2\mathbb{Z}$ sends the $SO(3)$-invariant 2-form $\omega$ to minus itself. The physics action in this case can be written using Witten's construction as the integral of the volume form over a 2-chain $b$ (representing a disk) bounding the 1-cycle representing the worldline. The shift in the action may be written as $\int_b (L^*_{gG_0} -1)\omega = -2 \int_b \omega$, which must equal an integer. Shrinking the worldline and disk to a point shows that it must equal zero, and hence $\omega =0$.

We will return to the issue of the general classification of both AB and WZ terms for disconected $G$ in \cite{JDBGORW}.

\bibliography{articles_bib}
\bibliographystyle{JHEP}
\end{document}